\documentclass[%
 reprint,
 superscriptaddress,
 amsmath,amssymb,
 aps,
 prl,
]{revtex4-1}

\usepackage{graphicx}
\usepackage{dcolumn}
\usepackage{bm}
\usepackage{booktabs}
\usepackage{siunitx}
\usepackage[hidelinks]{hyperref}
\usepackage[utf8]{inputenc}
\usepackage{wrapfig}

\begin{document}

\preprint{APS/123-QED}

\title{A compact and simple photonic nano-antenna for efficient photon extraction from nitrogen vacancy centers in bulk diamond}

\author{Niko Nikolay}
\author{Günter Kewes}%
\author{Oliver Benson}%
\affiliation{%
 Institut für Physik, Humboldt-Universität zu Berlin,\\
 Newtonstraße 15, D-12489 Berlin, Germany
}%





\begin{abstract}
Enhancing the collection efficiency of single emitters via dielectric antennas is a common approach in research dealing with solid-state quantum light sources. Current design concepts often suffer from a large footprint, quenching or a complex fabrication process. Here we report on another design where the Kerker effect is used to collimate emission from a nitrogen vacancy (NV) center. The structure discussed here is an easy to fabricate cylindrical protrusion made from diamond on top of a diamond substrate with NVs incorporated into its center. We optimize the structure with respect to application in quantum information processing and quantum sensing and find an up to 15 times higher photon extraction.

\end{abstract}

\maketitle


\section{\label{sec:Introduction}Introduction}

In recent years promising concepts for quantum sensing (QS) and quantum information processing (QIP) have been proposed and demonstrated, which desire single photons of differing "quality" \cite{Maletinsky2012,Schell2014,Dutt2007,Barrett2005}. While high single photon fluxes are desired without the need for indistinguishability for QS, QIP relies on the exchange of quantum states between isolated quantum bits (qubits) and on the quantum interference of those states \cite{Duan2010,Somaschi2016}.

Intensively studied single photon emitters are, e.g., molecules \cite{Lee2011}, quantum dots \cite{Buckley2012}, or defects in hBN crystals \cite{Tran2015} or diamond, like the silicon vacancy (SiV) \cite{Neu2011,Sipahigil2016} and nitrogen vacancy (NV) center \cite{Jelezko2006}; all of them feature some pros and cons when compared to each other. NVs however seem to offer a high potential for the implementation in both, QS and QIP. NVs feature long spin coherence times and their energy splitting of \SI{2.87}{\giga\hertz} is compatible to well established and commercially available technologies, thus allowing for easy implementation of optically detected magnetic resonance (ODMR) experiments \cite{Gruber1997}.

Apart from the sole properties of the single photon emitter it is of equal importance to consider ways of integration into practical devices. In this regard it is important to mention, that NV defect centers can be positioned by various methods with high lateral and vertical precision of around \SI{30}{\nano\meter} into diamond nowadays \cite{Meijer2008,Pezzagna2011}. With respect to QS and QIP routes for efficient photon extraction are needed, e.g., exploiting micro- and nano-antenna structures, as the collection efficiency from bulk diamond is often limited by total internal reflection and trapping inside the host material of the emitter. 

\begin{figure}
 \includegraphics[width=\columnwidth]{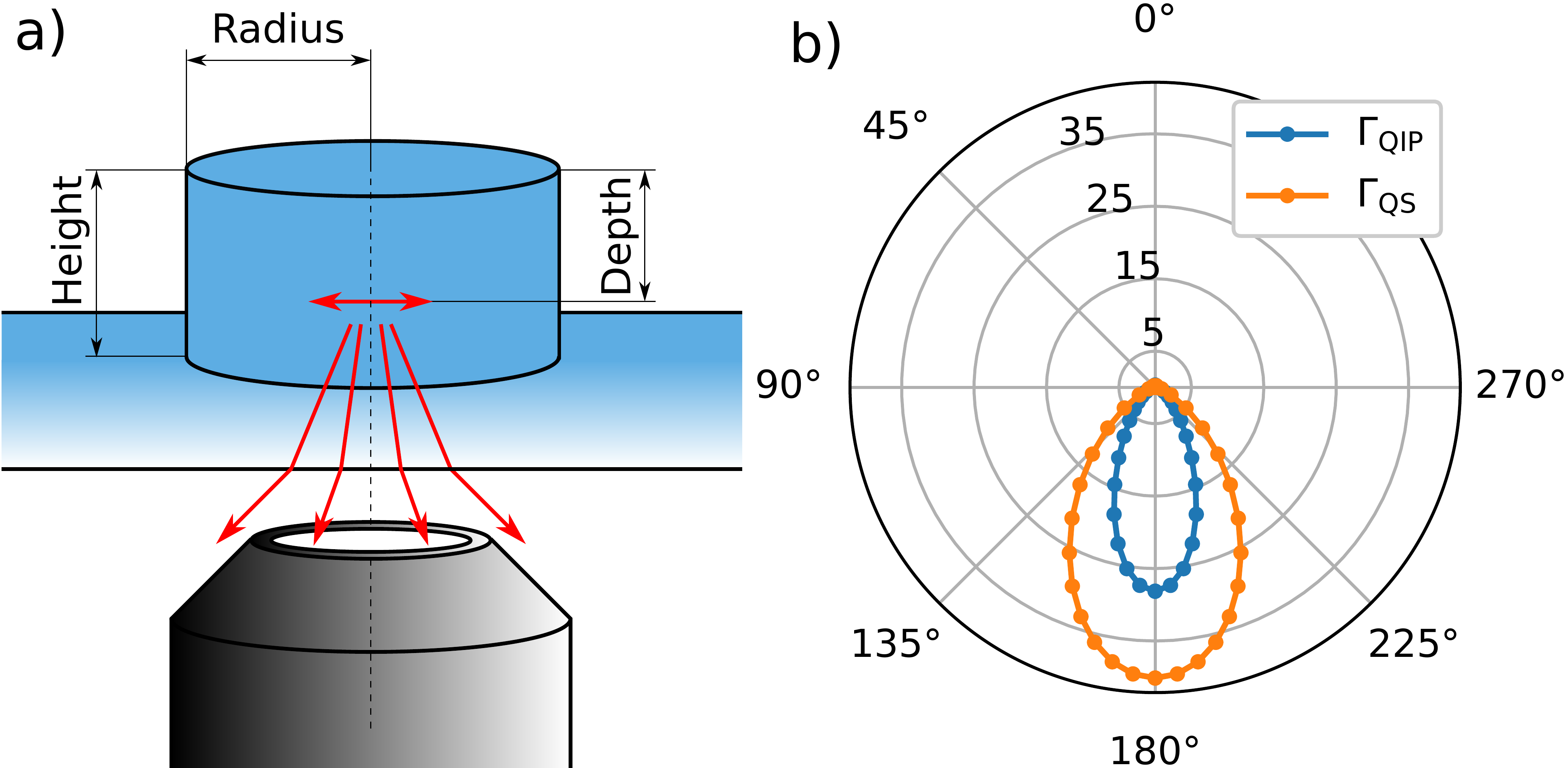}
 \caption{\textit{Antenna schematics and far field emission pattern.} In a) the antenna is shown schematically with three geometrical values. An oscillation dipole is denoted by a red double arrow within the structure. Emission from this dipole is collimated by the antenna and collected via an objective lens. b) shows a polar plot of the collimated far field emission caused by optimized antennas for QIP and QS.}
 \label{fig:schematics}
\end{figure}

A huge variety of micro- and nano-antennas have been discussed in literature \cite{Rajasekharan2015,Novotny2011,Muhlschlegel2005,Agio2012}. Those structures are crucial for the success of certain experiments, e.g. for entanglement measurements with NVs, solid immersion lenses (SILs) were used \cite{Hensen2015}. For a practical implementation the antenna footprint, its efficiency as well as its simplicity are crucial criteria, in particular for an on chip integration to build complex networks or cost efficient sensors. The antennas featuring the smallest footprint can be found in plasmonics but also purely dielectric antennas with fully sub-micron dimensions were introduced recently, based on high refractive indexes \cite{Fu2013}. It was proposed to make use of the Kerker effect to control the emission directivity of nearby emitters \cite{Rolly2012,Staude2013} - such a system represents a hybrid system of antenna with emitter \cite{Benson2011}.

Here we investigate a simpler modified approach where we consider an emitter \textit{inside} the nano-antenna of moderate refractive index, i.e., a compact photonic nano-antenna which is based on the Kerker effect \cite{Kerker1983}. We refer to this as a Kerker-like effect, since we are not exciting the antenna from a specific side but from within. The antenna yields a moderate enhancement of the radiative emission rate accompanied by a boost in emission directivity. We focus on NVs in bulk diamond, though the physical mechanism behind the Kerker effect is fundamental and can be utilized to any emitter and any dielectric host medium in principle. The antenna structure can be realized by a simple state-of-the-art structuring of the diamond surface producing cylindrical protrusions without the need of additional materials. 

With respect to applications in QS we design the antenna to work with shallow NV centers, i.e., the defect is closely positioned to the surface (we restrict the implantation depth to be between \SIlist{4;50}{\nano\meter}). To this end we optimize the nano antenna to collect the entire NV spectrum ranging roughly from \SIrange{600}{800}{\nano\meter} to harvest as many photons as possible. Typically ODMR sequences are used for QS, in this case the photon detection rates determine the signal to noise ratio of the measurements and by this the needed acquisition times. 
For QIP we optimize the antenna for collecting zero-phonon-line (ZPL) light only. The optimization was performed under the restriction of having a dipole implantation depth of more than \SI{50}{\nano\meter} to shield the NV against the environment, preserving the superior properties known from bulk samples, like ZPL-emission with low spectral diffusion and long coherence times of the NV ground state \cite{Ohno2012}.

\autoref{fig:schematics} a) shows the schematics of the proposed antenna design: A cylindrical protrusion etched into bulk diamond with an embedded NV. The idea behind that design can be explained as follows: The NV couples to localized electric and magnetic modes (analog to the "natural modes" in Mie theory) which are supported by the disc-shaped antenna due to the geometry and the relatively high refractive index of diamond (n $\sim 2.4$). These localized modes then couple to radiative modes in the diamond substrate, i.e., the localized modes are not "dark" but "bright" or "leaky". The spectral position of each mode is determined by antenna height and radius. Furthermore the coupling efficiency of the NV with the modes depends on the NV position and orientation within the structure. Here we focus on a vertical orientation of the NV’s defect axis since this is the case for widely used (111) bulk diamond \cite{Michl2014}. Tuning height and radius of the cylindrical protrusion to spectrally match magnetic and an electric modes lead to the ordinary Kerker effect or a higher order Kerker effect \cite{Alaee2015,Kerker1983}, as we will demonstrate below. In such a case the NV emission gets directed into the substrate with a narrow angular distribution and its lifetime will get shortened due to the Purcell effect.

\section{\label{sec:Simulation}Simulation}

In order to optimize and analyze the proposed antenna we define a parameter $\Gamma$, which quantifies the figure of merit, i.e. the envisioned performance most concisely. This $\Gamma$-Parameter describes the normalized rate of photons that can be harvested, i.e., which can be actually collected by the optical setup (characterized by a numerical aperture (NA)). The normalization is performed with respect to a reference structure; here a NV near the diamond-air surface at an optimal implantation depth (QIP \SI{140}{\nano\meter}, QS \SI{150}{\nano\meter}) is used for this purpose, as such structures are frequently used (further details can be found in the supplementary material). This reference is used for any calculation shown in the following. For each realization of the antenna with an NV center the corresponding lifetime $\tau$ (or inverse emission rate) and the collection efficiency $\eta_\text{C}$ are calculated. $\eta_\text{C}$ is a function of the NA, but also the angle dependent Fresnel-transmissions at the diamond-air surface at the back side of the bulk diamond is considered. Both, $\tau$ and $\eta_C$, are combined in the following manner:

\begin{equation}
\label{eq:FOM}
\Gamma = \frac{\tau'}{\tau}\frac{\int\limits_{\Delta\lambda}\rho(\lambda)\,\eta_C(NA,\lambda)\,\mathbf{d}\lambda}{\int\limits_{\Delta\lambda}\rho(\lambda)\,\eta'_C(NA,\lambda)\,\mathbf{d}\lambda}
\end{equation}

This number displays the total photon count enhancement at a desired wavelength range $\Delta\lambda$. As the collection efficiency is also a function of wavelength, it is weighted with the spectral density of the NV-spectrum $\rho(\lambda)$ and integrated over the region of interest $\Delta\lambda$. For a systematic characterization of the lifetime shortening and the collection efficiency, a finite-element solver is used (JCMwave). Further details of this computation are given in the supplement. Values denoted by $'$ are reference variables. 

In a first step the geometric parameters of the antenna are optimized by maximizing $\Gamma$. To this end we used the Nelder-Mead method \cite{Nelder1965} and three free parameters: disk height, disk radius and NV center implantation depth. With respect to QIP $\Delta\lambda$ is chosen to cover the zero phonon line (ZPL) occurring at \SI{637}{\nano\meter} (shown in the supplement), for QS $\Delta\lambda$ is chosen to cover the entire NV-spectrum. Optimized geometrical values are shown in \autoref{tab:results}.

\begin{table}[h]
 \caption{\textit{Optimized values.} Optimal geometric dimensions for each $\Gamma$ are shown for different collection NAs. Geometric dimensions are given in nm.}
 \begin{ruledtabular}
  \begin{tabular}{c c c c c c c c}
   & height& radius & depth & $\Gamma^{0.4}$ & $\Gamma^{0.6}$ & $\Gamma^{0.9}$ & $\Gamma^{1.4, \text{oil}}$\\
   \hline
   QIP	& $579$ & $321$ & $141$ & $15.4$ & $11.0$ & $7.8$ & $3.1$ \\
   QS	& $359$ & $261$  & $5.5$ & $15.3$ & $11.3$ & $6.5$ & $2.9$ \\
  \end{tabular}
 \end{ruledtabular}
 \label{tab:results}
\end{table}

\section{\label{sec:Analyzing}Analyzing structure variations and functionality}

\begin{figure}[h]
 \includegraphics[width=\columnwidth]{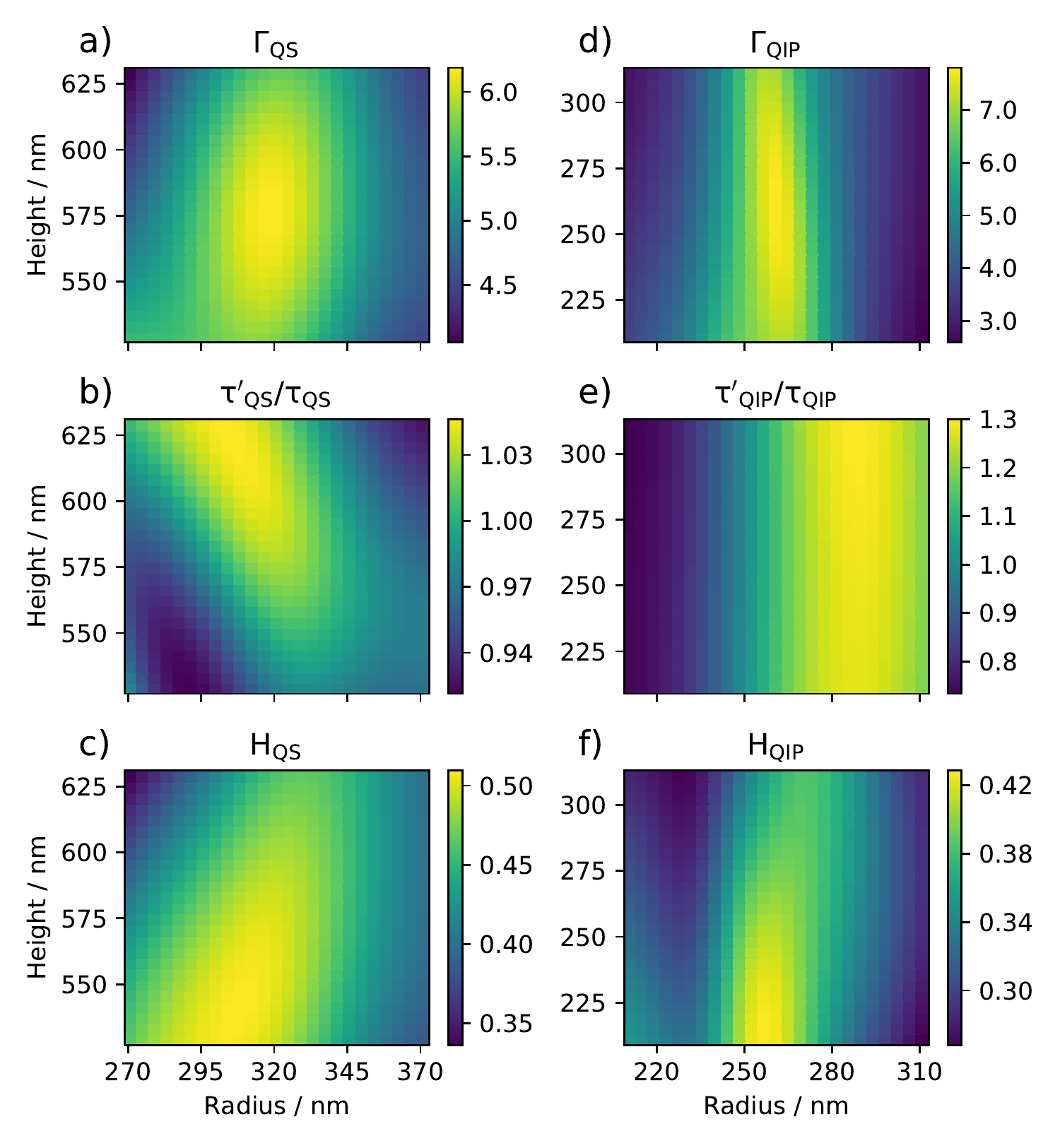}%
 \caption{\textit{$\varGamma$-parameters and its parts scanned over geometrical parameters.} $\Gamma$-parameters (a and d), the corresponding relative reduction of lifetimes $\tau'/\tau$ (b and e) and weighted collection efficiencies $H$ (c and f) are shown. Figures a), b), c) and d), e), f) share the same x axis respectively. The product of b) and c) gives a) - and the product of e) and f) gives d) up to a normalization by $H'$.}
 \label{fig:heatmaps}
\end{figure}

In the following we analyze the derived optimum values for $\Gamma$ in more detail, i.e., we investigate their sensitivity to geometrical changes and study which physical effects lead to the improvement. We focus on a fixed numerical aperture of 0.9 for simplicity. Therefore each $\Gamma$ is computed for sets of heights and radii, shown in \autoref{fig:heatmaps}. The $\Gamma$-parameter consists of the collection efficiency and a change in lifetime, which in turn is connected to a modification of the spectrum of the emitter. Besides $\Gamma$, the corresponding relative reduction of the lifetime $\tau'/\tau$ and the spectrally weighted collection efficiency $H$, which is the denominator of \autoref{eq:FOM} excluding $\tau'$ is shown in \autoref{fig:heatmaps} for each $\Gamma$ correspondingly. $H$ gives the percentage of collected photons over the whole spectrum.

Highest $\Gamma$s are reached when the weighted collection efficiency and the lifetime shortening show high values at same radius and height. The antenna optimized for $\Gamma_\text{QS}$ is less sensitive to geometrical changes compared to the one optimized for $\Gamma_\text{QIP}$. However, changes in disk size only weakly affect the antenna performance.

\subsection{\label{sec:NAdependence}NA dependence of optimum structure}

\begin{figure}
 \includegraphics[width=\columnwidth]{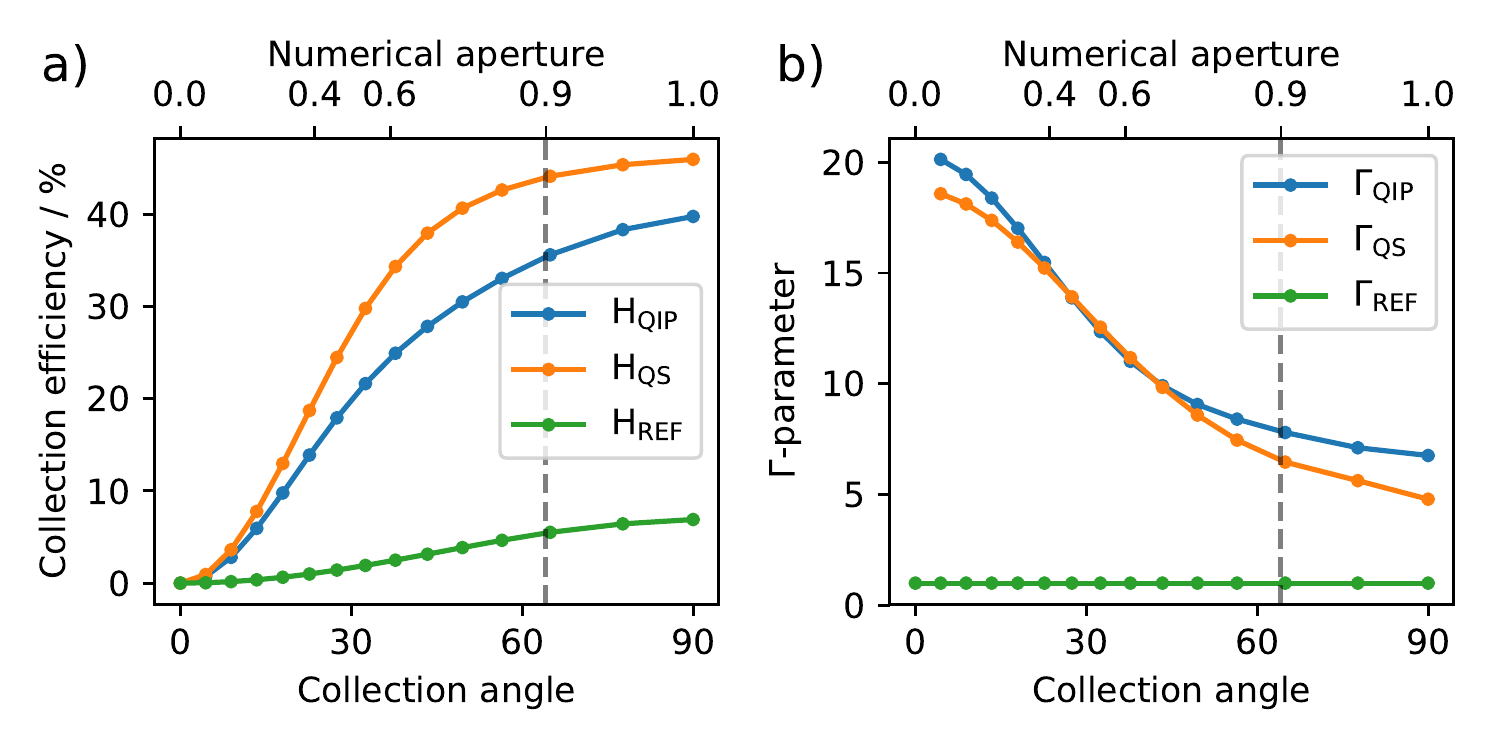}%
 \caption{\textit{Detection angle dependent collection efficiency and $\varGamma$-parameters.} With optimal geometric parameters the collection efficiency (a) and $\Gamma_\text{QS/QIP/REF}$ (b) are shown with respect to the collection angle. A dashed gray line indicates an NA of 0.9.}
 \label{fig:nadependence}
\end{figure}
As can bee seen in \autoref{fig:heatmaps}, the antenna alters the dipole emission and leads to high directivities that appear only within a small range of angles. While we only quoted the $\Gamma_\text{QS}$ and $\Gamma_\text{QIP}$ for certain NAs in \autoref{tab:results} we show a detailed analysis of how $\Gamma$ and collection efficiency depend on the NA in \autoref{fig:nadependence} a). Collection efficiencies with respect to the collection angle/NA are shown for parmaters $\Gamma_\text{QS}$ and $\Gamma_\text{QIP}$ as well as for the reference structure. Compared to the reference structure, a dramatic enhancement in collection efficiency occurs. While $\Gamma_\text{QS}$ increases mostly by raising collection efficiency and lifetime shortening, $\Gamma_\text{QIP}$ also profits from a spectral altering of the emission spectrum. For this reason \autoref{fig:nadependence} a) shows a lower collection efficiency for QIP compared with QS while in b) QIP reaches mostly the same values.

\subsection{\label{sec:displacement}Lateral and vertical dipole displacement}

\begin{figure}
 \includegraphics[width=\columnwidth]{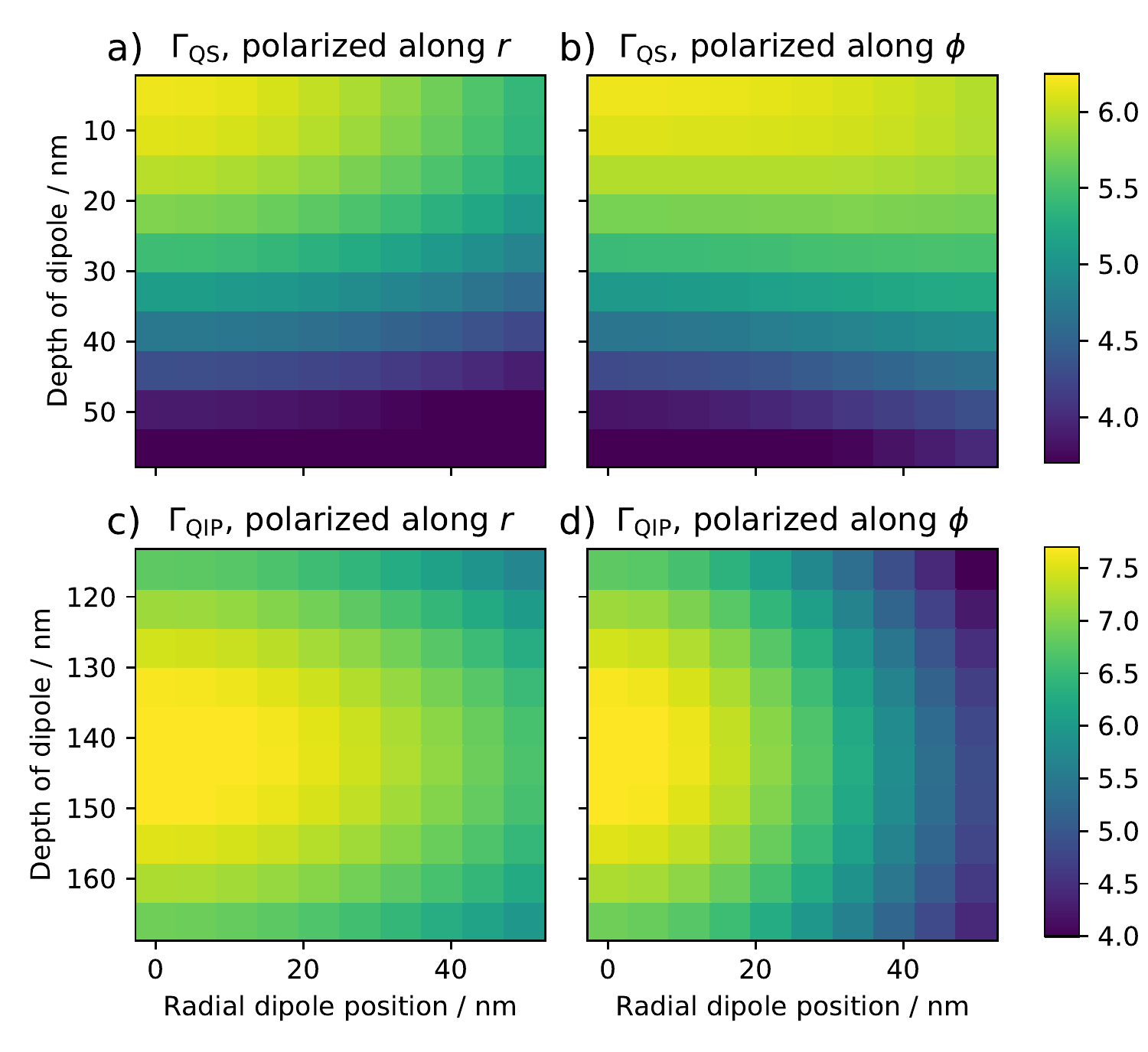}%
 \caption{\textit{The effect on $\varGamma$-parameters when displacing the dipole.} $\Gamma_\text{QS}$ in a) and b) and $\Gamma_\text{QIP}$ in c) d) are shown with respect to the radial dipole position and dipole depth with optimal geometric parameters, respectively. The dipole polarization is chosen to be along the $r$ component of a vector in cylindrical coordinates in a) and c) and along the $\phi$ component in b) and d).}
 \label{fig:displacement}
\end{figure}

Until now a perfect alignment relative to the optimal antenna structure was assumed, i.e. a position on the symmetry axis of the cylinder at the optimal depth. In a realistic experiment the NV could be implanted after fabricating the antenna structure. Today's implantation techniques reach lateral resolutions down to \SI{30}{\nano\meter} [29] and a depth resolution of up to $\SI{50}{\nano\meter}$ for depths mentioned here \cite{Pezzagna2011}. \autoref{fig:displacement} shows the influence on $\Gamma$-parameters when the dipole is misaligned within that range both in lateral and vertical direction as a heat map.

The NV has two optical transitions, each corresponds to a linear transition dipole moment. Both dipolar transitions are perpendicular to each other and perpendicular to the NV axis. All simulations done before take this into account automatically since the antenna-dipole-configuration is rotationally symmetric when a dipole is positioned exactly on the symmetry axis. However, under radial displacement the symmetry is broken and its orientation becomes more important. To take this into account, simulations with a dipole displacement were done with two distinct dipole orientations. In one case the dipole is oriented along the $r$ component, in the other case it is oriented along the $\phi$ component of a vector in cylindrical coordinates, although an arbitrary dipole orientation in between is not necessarily just a simple superposition of both cases. However, it seems to be sufficient to examine those two polarizations to qualitatively evaluate the antenna performance under displacement.

A displacement in the implantation depth has only a moderate impact on $\Gamma$-parameter (at least in this approximate range of reported fabrication precision). Displacing the defect laterally by, e.g., \SI{15}{\nano\meter}, which corresponds to the discussed resolution limit, causes only a slight change in performance. Moving the dipole position and tilting its orientation changes the coupling efficiency between emitter and eigenmodes of the antenna. This leads to less constructive interference into observable angles and thus to a lower $\Gamma$-parameter. However, we conclude that the antenna's $\Gamma$-parameter is rather robust against implantation misalignments.

\section{\label{sec:Kerker}Kerker effect}

\begin{figure}
 \includegraphics[width=\columnwidth]{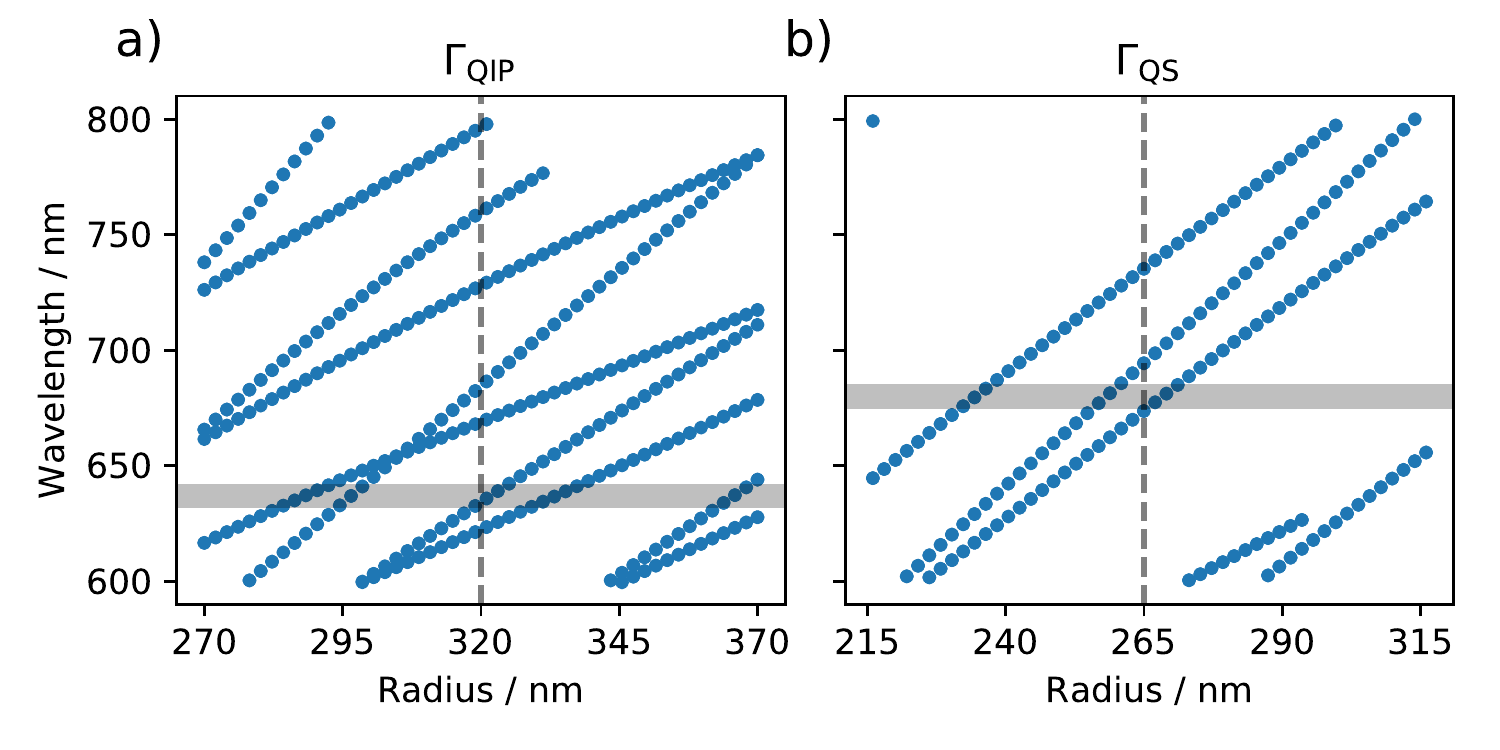}%
 \caption{\textit{Radius and wavelength dependence of resonances provided by the antenna structure.} Blue dots denote the central wavelength of different eigenmodes hosted by the antenna. The height is fixed in a) to \SI{579}{\nano\meter} and \SI{359}{\nano\meter} in b). The vertical dashed gray line denotes the optimal radius, the horizontal gray line indicates for the design wavelength for $\Gamma_\text{QIP}$ in a) and for the peak emission of the NV at room temperature in b). Neighboring modes can be obtained which are able to interfere.}
 \label{fig:resonances}
\end{figure}

We identify the Kerker effect or a Kerker-like effect to be the physical mechanism behind the collimation effect. The Kerker effect arises when two localized modes oscillate with a phase difference which results in regions of constructive/destructive interference in the far-field. Kerker was the first who recognized, that the interference of a magnetic and an electric dipole mode yield a pronounced directivety of the emitted photon flux\cite{Kerker1983}. Also electric dipole and quadrupole modes were used to obtain the Kerker effect, which can also be achieved with other combinations of higher order modes \cite{Alaee2015,Person2013}. \autoref{fig:resonances} shows the spectral evolution of modes for both $\Gamma_\text{QS}$ and $\Gamma_\text{QIP}$. While the height is fixed at the values taken from \autoref{tab:results}, respectively, the antenna radius is varied. At the optimal radius and ZPL-wavelength, indicated by the line crossing (vertical and horizontal gray lines) in \autoref{fig:resonances} a), two modes are close to each other. This indicates that a higher order Kerker condition is fulfilled, which is known to occur slightly shifted with respect to the crossing point \cite{Alaee2015,Kerker1983}. \autoref{fig:resonances} b) shows the modes in the antenna that is optimized for QS. Even though we used the whole spectrum of the NV for the optimization, the optimal structure shows two resonances at \SI{680}{\nano\meter}, i.e., where the NV spectrum has its maximum (see supplementary material). We conclude that the higher order Kerker effect occurs at this wavelength and maximizes $\Gamma_\text{QS}$.

A destructive interference in the upper plane and constructive interference in the lower plain of those two modes lead to an enhanced collection efficiency which is one ingredient to maximize the $\Gamma$-parameter.

By changing its geometrical dimensions it can be tuned to enhance emission of any wavelength, in forward or backward direction. Even other materials could be used such as GaAs with quantum dots incorporated into the antenna emitting into free space rather than into the non transparent substrate.

\section{\label{sec:Conclusion}Conclusion}

In conclusion we presented an extremely versatile antenna design characterized with an experimentally relevant figure of merit for collimating antennas. We consider two scenarios in which either spectrally broad or narrow-band emitters are of interest, i.e., either for their implementation into QIP or QS technology. The $\Gamma$-parameter considers the effective reduction of the emitter's lifetime, a changed collection efficiency and a spectral alteration of the emitter emission spectrum. With this $\Gamma$-parameter we were able to find optimal geometrical parameters of disks made from diamond on top of a diamond substrate with an integrated nitrogen vacancy center. In one of the two analyzed configurations the NV will preserve its spin coherence time while the photon flux stemming from the ZPL is maximized. The other configuration maximizes the whole photon flux from shallow NVs used for sensing applications. The presence of the proposed antenna show an intensity enhancement of 7.8 (QIP) and 6.6 (QS) even for a rather high NA of 0.9. A big advantage of this antenna type is the efficient detection on the bulk diamond side. Compared to bulls eye antennas of various types (made from metal and/or dielectrics) the single element cylindrical protrusion is simple to fabricate, small in footprint and avoids quenching as known from plasmonic antennas \cite{Livneh2016,Li2015,Moroz2010}. An eigenmode analysis uncovers the working principle of this devices, which turns out to be the Kerker effect. Large parameter scans shows the robustness of this design with respect to geometrical imperfections. Simulations with displaced dipoles within experimentally demonstrated implantation resolutions show just a moderate lowering of the $\Gamma$-parameter. While we focused in this work on NVs in diamond, this approach can be applied to basically any emitter and host material like widely used quantum dots integrated in GaAs chips.

The authors thank F. Böhm, N. Sadzak and B. Sontheimer for fruitful discussions. We acknowledge financial support by the Einstein Foundation Berlin (ActiPlAnt) and BMBF (Nano-Film).

\bibliography{references}

\begin{thebibliography}{33}%
\makeatletter
\providecommand \@ifxundefined [1]{%
 \@ifx{#1\undefined}
}%
\providecommand \@ifnum [1]{%
 \ifnum #1\expandafter \@firstoftwo
 \else \expandafter \@secondoftwo
 \fi
}%
\providecommand \@ifx [1]{%
 \ifx #1\expandafter \@firstoftwo
 \else \expandafter \@secondoftwo
 \fi
}%
\providecommand \natexlab [1]{#1}%
\providecommand \enquote  [1]{``#1''}%
\providecommand \bibnamefont  [1]{#1}%
\providecommand \bibfnamefont [1]{#1}%
\providecommand \citenamefont [1]{#1}%
\providecommand \href@noop [0]{\@secondoftwo}%
\providecommand \href [0]{\begingroup \@sanitize@url \@href}%
\providecommand \@href[1]{\@@startlink{#1}\@@href}%
\providecommand \@@href[1]{\endgroup#1\@@endlink}%
\providecommand \@sanitize@url [0]{\catcode `\\12\catcode `\$12\catcode
  `\&12\catcode `\#12\catcode `\^12\catcode `\_12\catcode `\%12\relax}%
\providecommand \@@startlink[1]{}%
\providecommand \@@endlink[0]{}%
\providecommand \url  [0]{\begingroup\@sanitize@url \@url }%
\providecommand \@url [1]{\endgroup\@href {#1}{\urlprefix }}%
\providecommand \urlprefix  [0]{URL }%
\providecommand \Eprint [0]{\href }%
\providecommand \doibase [0]{http://dx.doi.org/}%
\providecommand \selectlanguage [0]{\@gobble}%
\providecommand \bibinfo  [0]{\@secondoftwo}%
\providecommand \bibfield  [0]{\@secondoftwo}%
\providecommand \translation [1]{[#1]}%
\providecommand \BibitemOpen [0]{}%
\providecommand \bibitemStop [0]{}%
\providecommand \bibitemNoStop [0]{.\EOS\space}%
\providecommand \EOS [0]{\spacefactor3000\relax}%
\providecommand \BibitemShut  [1]{\csname bibitem#1\endcsname}%
\let\auto@bib@innerbib\@empty
\bibitem [{\citenamefont {Maletinsky}\ \emph {et~al.}(2012)\citenamefont
  {Maletinsky}, \citenamefont {Hong}, \citenamefont {Grinolds}, \citenamefont
  {Hausmann}, \citenamefont {Lukin}, \citenamefont {Walsworth}, \citenamefont
  {Loncar},\ and\ \citenamefont {Yacoby}}]{Maletinsky2012}%
  \BibitemOpen
  \bibfield  {author} {\bibinfo {author} {\bibfnamefont {P.}~\bibnamefont
  {Maletinsky}}, \bibinfo {author} {\bibfnamefont {S.}~\bibnamefont {Hong}},
  \bibinfo {author} {\bibfnamefont {M.~S.}\ \bibnamefont {Grinolds}}, \bibinfo
  {author} {\bibfnamefont {B.}~\bibnamefont {Hausmann}}, \bibinfo {author}
  {\bibfnamefont {M.~D.}\ \bibnamefont {Lukin}}, \bibinfo {author}
  {\bibfnamefont {R.~L.}\ \bibnamefont {Walsworth}}, \bibinfo {author}
  {\bibfnamefont {M.}~\bibnamefont {Loncar}}, \ and\ \bibinfo {author}
  {\bibfnamefont {A.}~\bibnamefont {Yacoby}},\ }\href {\doibase
  10.1038/nnano.2012.50} {\bibfield  {journal} {\bibinfo  {journal} {Nature
  Nanotechnology}\ }\textbf {\bibinfo {volume} {7}},\ \bibinfo {pages} {320}
  (\bibinfo {year} {2012})}\BibitemShut {NoStop}%
\bibitem [{\citenamefont {Schell}\ \emph {et~al.}(2014)\citenamefont {Schell},
  \citenamefont {Engel}, \citenamefont {Werra}, \citenamefont {Wolff},
  \citenamefont {Busch},\ and\ \citenamefont {Benson}}]{Schell2014}%
  \BibitemOpen
  \bibfield  {author} {\bibinfo {author} {\bibfnamefont {A.~W.}\ \bibnamefont
  {Schell}}, \bibinfo {author} {\bibfnamefont {P.}~\bibnamefont {Engel}},
  \bibinfo {author} {\bibfnamefont {J.~F.~M.}\ \bibnamefont {Werra}}, \bibinfo
  {author} {\bibfnamefont {C.}~\bibnamefont {Wolff}}, \bibinfo {author}
  {\bibfnamefont {K.}~\bibnamefont {Busch}}, \ and\ \bibinfo {author}
  {\bibfnamefont {O.}~\bibnamefont {Benson}},\ }\href {\doibase
  10.1021/nl500460c} {\bibfield  {journal} {\bibinfo  {journal} {Nano Letters}\
  }\textbf {\bibinfo {volume} {14}},\ \bibinfo {pages} {2623} (\bibinfo {year}
  {2014})}\BibitemShut {NoStop}%
\bibitem [{\citenamefont {Dutt}\ \emph {et~al.}(2007)\citenamefont {Dutt},
  \citenamefont {Childress}, \citenamefont {Jiang}, \citenamefont {Togan},
  \citenamefont {Maze}, \citenamefont {Jelezko}, \citenamefont {Zibrov},
  \citenamefont {Hemmer},\ and\ \citenamefont {Lukin}}]{Dutt2007}%
  \BibitemOpen
  \bibfield  {author} {\bibinfo {author} {\bibfnamefont {M.~V.~G.}\
  \bibnamefont {Dutt}}, \bibinfo {author} {\bibfnamefont {L.}~\bibnamefont
  {Childress}}, \bibinfo {author} {\bibfnamefont {L.}~\bibnamefont {Jiang}},
  \bibinfo {author} {\bibfnamefont {E.}~\bibnamefont {Togan}}, \bibinfo
  {author} {\bibfnamefont {J.}~\bibnamefont {Maze}}, \bibinfo {author}
  {\bibfnamefont {F.}~\bibnamefont {Jelezko}}, \bibinfo {author} {\bibfnamefont
  {A.~S.}\ \bibnamefont {Zibrov}}, \bibinfo {author} {\bibfnamefont {P.~R.}\
  \bibnamefont {Hemmer}}, \ and\ \bibinfo {author} {\bibfnamefont {M.~D.}\
  \bibnamefont {Lukin}},\ }\href {\doibase 10.1126/science.1139831} {\bibfield
  {journal} {\bibinfo  {journal} {Science}\ }\textbf {\bibinfo {volume}
  {316}},\ \bibinfo {pages} {1312} (\bibinfo {year} {2007})}\BibitemShut
  {NoStop}%
\bibitem [{\citenamefont {Barrett}\ and\ \citenamefont
  {Kok}(2005)}]{Barrett2005}%
  \BibitemOpen
  \bibfield  {author} {\bibinfo {author} {\bibfnamefont {S.~D.}\ \bibnamefont
  {Barrett}}\ and\ \bibinfo {author} {\bibfnamefont {P.}~\bibnamefont {Kok}},\
  }\href {\doibase 10.1103/PhysRevA.71.060310} {\bibfield  {journal} {\bibinfo
  {journal} {Physical Review A}\ }\textbf {\bibinfo {volume} {71}},\ \bibinfo
  {pages} {060310} (\bibinfo {year} {2005})}\BibitemShut {NoStop}%
\bibitem [{\citenamefont {Duan}\ and\ \citenamefont {Monroe}(2010)}]{Duan2010}%
  \BibitemOpen
  \bibfield  {author} {\bibinfo {author} {\bibfnamefont {L.-M.}\ \bibnamefont
  {Duan}}\ and\ \bibinfo {author} {\bibfnamefont {C.}~\bibnamefont {Monroe}},\
  }\href {\doibase 10.1103/RevModPhys.82.1209} {\bibfield  {journal} {\bibinfo
  {journal} {Reviews of Modern Physics}\ }\textbf {\bibinfo {volume} {82}},\
  \bibinfo {pages} {1209} (\bibinfo {year} {2010})}\BibitemShut {NoStop}%
\bibitem [{\citenamefont {Somaschi}\ \emph {et~al.}(2016)\citenamefont
  {Somaschi}, \citenamefont {Giesz}, \citenamefont {{De Santis}}, \citenamefont
  {Loredo}, \citenamefont {Almeida}, \citenamefont {Hornecker}, \citenamefont
  {Portalupi}, \citenamefont {Grange}, \citenamefont {Ant{\'{o}}n},
  \citenamefont {Demory}, \citenamefont {G{\'{o}}mez}, \citenamefont {Sagnes},
  \citenamefont {Lanzillotti-Kimura}, \citenamefont {Lema{\'{i}}tre},
  \citenamefont {Auffeves}, \citenamefont {White}, \citenamefont {Lanco},\ and\
  \citenamefont {Senellart}}]{Somaschi2016}%
  \BibitemOpen
  \bibfield  {author} {\bibinfo {author} {\bibfnamefont {N.}~\bibnamefont
  {Somaschi}}, \bibinfo {author} {\bibfnamefont {V.}~\bibnamefont {Giesz}},
  \bibinfo {author} {\bibfnamefont {L.}~\bibnamefont {{De Santis}}}, \bibinfo
  {author} {\bibfnamefont {J.~C.}\ \bibnamefont {Loredo}}, \bibinfo {author}
  {\bibfnamefont {M.~P.}\ \bibnamefont {Almeida}}, \bibinfo {author}
  {\bibfnamefont {G.}~\bibnamefont {Hornecker}}, \bibinfo {author}
  {\bibfnamefont {S.~L.}\ \bibnamefont {Portalupi}}, \bibinfo {author}
  {\bibfnamefont {T.}~\bibnamefont {Grange}}, \bibinfo {author} {\bibfnamefont
  {C.}~\bibnamefont {Ant{\'{o}}n}}, \bibinfo {author} {\bibfnamefont
  {J.}~\bibnamefont {Demory}}, \bibinfo {author} {\bibfnamefont
  {C.}~\bibnamefont {G{\'{o}}mez}}, \bibinfo {author} {\bibfnamefont
  {I.}~\bibnamefont {Sagnes}}, \bibinfo {author} {\bibfnamefont {N.~D.}\
  \bibnamefont {Lanzillotti-Kimura}}, \bibinfo {author} {\bibfnamefont
  {A.}~\bibnamefont {Lema{\'{i}}tre}}, \bibinfo {author} {\bibfnamefont
  {A.}~\bibnamefont {Auffeves}}, \bibinfo {author} {\bibfnamefont {A.~G.}\
  \bibnamefont {White}}, \bibinfo {author} {\bibfnamefont {L.}~\bibnamefont
  {Lanco}}, \ and\ \bibinfo {author} {\bibfnamefont {P.}~\bibnamefont
  {Senellart}},\ }\href {\doibase 10.1038/nphoton.2016.23} {\bibfield
  {journal} {\bibinfo  {journal} {Nature Photonics}\ }\textbf {\bibinfo
  {volume} {10}},\ \bibinfo {pages} {340} (\bibinfo {year} {2016})}\BibitemShut
  {NoStop}%
\bibitem [{\citenamefont {Lee}\ \emph {et~al.}(2011)\citenamefont {Lee},
  \citenamefont {Chen}, \citenamefont {Eghlidi}, \citenamefont {Kukura},
  \citenamefont {Lettow}, \citenamefont {Renn}, \citenamefont {Sandoghdar},\
  and\ \citenamefont {G{\"{o}}tzinger}}]{Lee2011}%
  \BibitemOpen
  \bibfield  {author} {\bibinfo {author} {\bibfnamefont {K.~G.}\ \bibnamefont
  {Lee}}, \bibinfo {author} {\bibfnamefont {X.~W.}\ \bibnamefont {Chen}},
  \bibinfo {author} {\bibfnamefont {H.}~\bibnamefont {Eghlidi}}, \bibinfo
  {author} {\bibfnamefont {P.}~\bibnamefont {Kukura}}, \bibinfo {author}
  {\bibfnamefont {R.}~\bibnamefont {Lettow}}, \bibinfo {author} {\bibfnamefont
  {A.}~\bibnamefont {Renn}}, \bibinfo {author} {\bibfnamefont {V.}~\bibnamefont
  {Sandoghdar}}, \ and\ \bibinfo {author} {\bibfnamefont {S.}~\bibnamefont
  {G{\"{o}}tzinger}},\ }\href {\doibase 10.1038/nphoton.2010.312} {\bibfield
  {journal} {\bibinfo  {journal} {Nature Photonics}\ }\textbf {\bibinfo
  {volume} {5}},\ \bibinfo {pages} {166} (\bibinfo {year} {2011})}\BibitemShut
  {NoStop}%
\bibitem [{\citenamefont {Buckley}\ \emph {et~al.}(2012)\citenamefont
  {Buckley}, \citenamefont {Rivoire},\ and\ \citenamefont
  {Vu{\v{c}}kovi{\'{c}}}}]{Buckley2012}%
  \BibitemOpen
  \bibfield  {author} {\bibinfo {author} {\bibfnamefont {S.}~\bibnamefont
  {Buckley}}, \bibinfo {author} {\bibfnamefont {K.}~\bibnamefont {Rivoire}}, \
  and\ \bibinfo {author} {\bibfnamefont {J.}~\bibnamefont
  {Vu{\v{c}}kovi{\'{c}}}},\ }\href {\doibase 10.1088/0034-4885/75/12/126503}
  {\bibfield  {journal} {\bibinfo  {journal} {Reports on Progress in Physics}\
  }\textbf {\bibinfo {volume} {75}},\ \bibinfo {pages} {126503} (\bibinfo
  {year} {2012})}\BibitemShut {NoStop}%
\bibitem [{\citenamefont {Tran}\ \emph {et~al.}(2015)\citenamefont {Tran},
  \citenamefont {Bray}, \citenamefont {Ford}, \citenamefont {Toth},\ and\
  \citenamefont {Aharonovich}}]{Tran2015}%
  \BibitemOpen
  \bibfield  {author} {\bibinfo {author} {\bibfnamefont {T.~T.}\ \bibnamefont
  {Tran}}, \bibinfo {author} {\bibfnamefont {K.}~\bibnamefont {Bray}}, \bibinfo
  {author} {\bibfnamefont {M.~J.}\ \bibnamefont {Ford}}, \bibinfo {author}
  {\bibfnamefont {M.}~\bibnamefont {Toth}}, \ and\ \bibinfo {author}
  {\bibfnamefont {I.}~\bibnamefont {Aharonovich}},\ }\href {\doibase
  10.1038/nnano.2015.242} {\bibfield  {journal} {\bibinfo  {journal} {Nature
  Nanotechnology}\ }\textbf {\bibinfo {volume} {11}},\ \bibinfo {pages} {37}
  (\bibinfo {year} {2015})}\BibitemShut {NoStop}%
\bibitem [{\citenamefont {Neu}\ \emph {et~al.}(2011)\citenamefont {Neu},
  \citenamefont {Steinmetz}, \citenamefont {Riedrich-M{\"{o}}ller},
  \citenamefont {Gsell}, \citenamefont {Fischer}, \citenamefont {Schreck},\
  and\ \citenamefont {Becher}}]{Neu2011}%
  \BibitemOpen
  \bibfield  {author} {\bibinfo {author} {\bibfnamefont {E.}~\bibnamefont
  {Neu}}, \bibinfo {author} {\bibfnamefont {D.}~\bibnamefont {Steinmetz}},
  \bibinfo {author} {\bibfnamefont {J.}~\bibnamefont {Riedrich-M{\"{o}}ller}},
  \bibinfo {author} {\bibfnamefont {S.}~\bibnamefont {Gsell}}, \bibinfo
  {author} {\bibfnamefont {M.}~\bibnamefont {Fischer}}, \bibinfo {author}
  {\bibfnamefont {M.}~\bibnamefont {Schreck}}, \ and\ \bibinfo {author}
  {\bibfnamefont {C.}~\bibnamefont {Becher}},\ }\href {\doibase
  10.1088/1367-2630/13/2/025012} {\bibfield  {journal} {\bibinfo  {journal}
  {New Journal of Physics}\ }\textbf {\bibinfo {volume} {13}},\ \bibinfo
  {pages} {025012} (\bibinfo {year} {2011})}\BibitemShut {NoStop}%
\bibitem [{\citenamefont {Sipahigil}\ \emph {et~al.}(2016)\citenamefont
  {Sipahigil}, \citenamefont {Evans}, \citenamefont {Sukachev}, \citenamefont
  {Burek}, \citenamefont {Borregaard}, \citenamefont {Bhaskar}, \citenamefont
  {Nguyen}, \citenamefont {Pacheco}, \citenamefont {Atikian}, \citenamefont
  {Meuwly}, \citenamefont {Camacho}, \citenamefont {Jelezko}, \citenamefont
  {Bielejec}, \citenamefont {Park}, \citenamefont {Lon{\v{c}}ar},\ and\
  \citenamefont {Lukin}}]{Sipahigil2016}%
  \BibitemOpen
  \bibfield  {author} {\bibinfo {author} {\bibfnamefont {A.}~\bibnamefont
  {Sipahigil}}, \bibinfo {author} {\bibfnamefont {R.~E.}\ \bibnamefont
  {Evans}}, \bibinfo {author} {\bibfnamefont {D.~D.}\ \bibnamefont {Sukachev}},
  \bibinfo {author} {\bibfnamefont {M.~J.}\ \bibnamefont {Burek}}, \bibinfo
  {author} {\bibfnamefont {J.}~\bibnamefont {Borregaard}}, \bibinfo {author}
  {\bibfnamefont {M.~K.}\ \bibnamefont {Bhaskar}}, \bibinfo {author}
  {\bibfnamefont {C.~T.}\ \bibnamefont {Nguyen}}, \bibinfo {author}
  {\bibfnamefont {J.~L.}\ \bibnamefont {Pacheco}}, \bibinfo {author}
  {\bibfnamefont {H.~A.}\ \bibnamefont {Atikian}}, \bibinfo {author}
  {\bibfnamefont {C.}~\bibnamefont {Meuwly}}, \bibinfo {author} {\bibfnamefont
  {R.~M.}\ \bibnamefont {Camacho}}, \bibinfo {author} {\bibfnamefont
  {F.}~\bibnamefont {Jelezko}}, \bibinfo {author} {\bibfnamefont
  {E.}~\bibnamefont {Bielejec}}, \bibinfo {author} {\bibfnamefont
  {H.}~\bibnamefont {Park}}, \bibinfo {author} {\bibfnamefont {M.}~\bibnamefont
  {Lon{\v{c}}ar}}, \ and\ \bibinfo {author} {\bibfnamefont {M.~D.}\
  \bibnamefont {Lukin}},\ }\href {\doibase 10.1126/science.aah6875} {\bibfield
  {journal} {\bibinfo  {journal} {Science}\ }\textbf {\bibinfo {volume}
  {354}},\ \bibinfo {pages} {847} (\bibinfo {year} {2016})}\BibitemShut
  {NoStop}%
\bibitem [{\citenamefont {Jelezko}\ and\ \citenamefont
  {Wrachtrup}(2006)}]{Jelezko2006}%
  \BibitemOpen
  \bibfield  {author} {\bibinfo {author} {\bibfnamefont {F.}~\bibnamefont
  {Jelezko}}\ and\ \bibinfo {author} {\bibfnamefont {J.}~\bibnamefont
  {Wrachtrup}},\ }\href {\doibase 10.1002/pssa.200671403} {\bibfield  {journal}
  {\bibinfo  {journal} {physica status solidi (a)}\ }\textbf {\bibinfo {volume}
  {203}},\ \bibinfo {pages} {3207} (\bibinfo {year} {2006})}\BibitemShut
  {NoStop}%
\bibitem [{\citenamefont {Gruber}(1997)}]{Gruber1997}%
  \BibitemOpen
  \bibfield  {author} {\bibinfo {author} {\bibfnamefont {A.}~\bibnamefont
  {Gruber}},\ }\href {\doibase 10.1126/science.276.5321.2012} {\bibfield
  {journal} {\bibinfo  {journal} {Science}\ }\textbf {\bibinfo {volume}
  {276}},\ \bibinfo {pages} {2012} (\bibinfo {year} {1997})}\BibitemShut
  {NoStop}%
\bibitem [{\citenamefont {Meijer}\ \emph {et~al.}(2008)\citenamefont {Meijer},
  \citenamefont {Pezzagna}, \citenamefont {Vogel}, \citenamefont {Burchard},
  \citenamefont {Bukow}, \citenamefont {Rangelow}, \citenamefont {Sarov},
  \citenamefont {Wiggers}, \citenamefont {Pl{\"{u}}mel}, \citenamefont
  {Jelezko}, \citenamefont {Wrachtrup}, \citenamefont {Schmidt-Kaler},
  \citenamefont {Schnitzler},\ and\ \citenamefont {Singer}}]{Meijer2008}%
  \BibitemOpen
  \bibfield  {author} {\bibinfo {author} {\bibfnamefont {J.}~\bibnamefont
  {Meijer}}, \bibinfo {author} {\bibfnamefont {S.}~\bibnamefont {Pezzagna}},
  \bibinfo {author} {\bibfnamefont {T.}~\bibnamefont {Vogel}}, \bibinfo
  {author} {\bibfnamefont {B.}~\bibnamefont {Burchard}}, \bibinfo {author}
  {\bibfnamefont {H.}~\bibnamefont {Bukow}}, \bibinfo {author} {\bibfnamefont
  {I.}~\bibnamefont {Rangelow}}, \bibinfo {author} {\bibfnamefont
  {Y.}~\bibnamefont {Sarov}}, \bibinfo {author} {\bibfnamefont
  {H.}~\bibnamefont {Wiggers}}, \bibinfo {author} {\bibfnamefont
  {I.}~\bibnamefont {Pl{\"{u}}mel}}, \bibinfo {author} {\bibfnamefont
  {F.}~\bibnamefont {Jelezko}}, \bibinfo {author} {\bibfnamefont
  {J.}~\bibnamefont {Wrachtrup}}, \bibinfo {author} {\bibfnamefont
  {F.}~\bibnamefont {Schmidt-Kaler}}, \bibinfo {author} {\bibfnamefont
  {W.}~\bibnamefont {Schnitzler}}, \ and\ \bibinfo {author} {\bibfnamefont
  {K.}~\bibnamefont {Singer}},\ }\href {\doibase 10.1007/s00339-008-4515-1}
  {\bibfield  {journal} {\bibinfo  {journal} {Applied Physics A}\ }\textbf
  {\bibinfo {volume} {91}},\ \bibinfo {pages} {567} (\bibinfo {year}
  {2008})}\BibitemShut {NoStop}%
\bibitem [{\citenamefont {Pezzagna}\ \emph {et~al.}(2011)\citenamefont
  {Pezzagna}, \citenamefont {Rogalla}, \citenamefont {Wildanger}, \citenamefont
  {Meijer},\ and\ \citenamefont {Zaitsev}}]{Pezzagna2011}%
  \BibitemOpen
  \bibfield  {author} {\bibinfo {author} {\bibfnamefont {S.}~\bibnamefont
  {Pezzagna}}, \bibinfo {author} {\bibfnamefont {D.}~\bibnamefont {Rogalla}},
  \bibinfo {author} {\bibfnamefont {D.}~\bibnamefont {Wildanger}}, \bibinfo
  {author} {\bibfnamefont {J.}~\bibnamefont {Meijer}}, \ and\ \bibinfo {author}
  {\bibfnamefont {A.}~\bibnamefont {Zaitsev}},\ }\href {\doibase
  10.1088/1367-2630/13/3/035024} {\bibfield  {journal} {\bibinfo  {journal}
  {New Journal of Physics}\ }\textbf {\bibinfo {volume} {13}},\ \bibinfo
  {pages} {035024} (\bibinfo {year} {2011})}\BibitemShut {NoStop}%
\bibitem [{\citenamefont {Rajasekharan}\ \emph {et~al.}(2015)\citenamefont
  {Rajasekharan}, \citenamefont {Kewes}, \citenamefont {Djalalian-Assl},
  \citenamefont {Ganesan}, \citenamefont {Tomljenovic-Hanic}, \citenamefont
  {McCallum}, \citenamefont {Roberts}, \citenamefont {Benson},\ and\
  \citenamefont {Prawer}}]{Rajasekharan2015}%
  \BibitemOpen
  \bibfield  {author} {\bibinfo {author} {\bibfnamefont {R.}~\bibnamefont
  {Rajasekharan}}, \bibinfo {author} {\bibfnamefont {G.}~\bibnamefont {Kewes}},
  \bibinfo {author} {\bibfnamefont {A.}~\bibnamefont {Djalalian-Assl}},
  \bibinfo {author} {\bibfnamefont {K.}~\bibnamefont {Ganesan}}, \bibinfo
  {author} {\bibfnamefont {S.}~\bibnamefont {Tomljenovic-Hanic}}, \bibinfo
  {author} {\bibfnamefont {J.~C.}\ \bibnamefont {McCallum}}, \bibinfo {author}
  {\bibfnamefont {A.}~\bibnamefont {Roberts}}, \bibinfo {author} {\bibfnamefont
  {O.}~\bibnamefont {Benson}}, \ and\ \bibinfo {author} {\bibfnamefont
  {S.}~\bibnamefont {Prawer}},\ }\href {\doibase 10.1038/srep12013} {\bibfield
  {journal} {\bibinfo  {journal} {Scientific Reports}\ }\textbf {\bibinfo
  {volume} {5}},\ \bibinfo {pages} {12013} (\bibinfo {year}
  {2015})}\BibitemShut {NoStop}%
\bibitem [{\citenamefont {Novotny}\ and\ \citenamefont {van
  Hulst}(2011)}]{Novotny2011}%
  \BibitemOpen
  \bibfield  {author} {\bibinfo {author} {\bibfnamefont {L.}~\bibnamefont
  {Novotny}}\ and\ \bibinfo {author} {\bibfnamefont {N.}~\bibnamefont {van
  Hulst}},\ }\href {\doibase 10.1038/nphoton.2010.237} {\bibfield  {journal}
  {\bibinfo  {journal} {Nature Photonics}\ }\textbf {\bibinfo {volume} {5}},\
  \bibinfo {pages} {83} (\bibinfo {year} {2011})}\BibitemShut {NoStop}%
\bibitem [{\citenamefont {Muhlschlegel}(2005)}]{Muhlschlegel2005}%
  \BibitemOpen
  \bibfield  {author} {\bibinfo {author} {\bibfnamefont {P.}~\bibnamefont
  {Muhlschlegel}},\ }\href {\doibase 10.1126/science.1111886} {\bibfield
  {journal} {\bibinfo  {journal} {Science}\ }\textbf {\bibinfo {volume}
  {308}},\ \bibinfo {pages} {1607} (\bibinfo {year} {2005})}\BibitemShut
  {NoStop}%
\bibitem [{\citenamefont {Agio}(2012)}]{Agio2012}%
  \BibitemOpen
  \bibfield  {author} {\bibinfo {author} {\bibfnamefont {M.}~\bibnamefont
  {Agio}},\ }\href {\doibase 10.1039/C1NR11116G} {\bibfield  {journal}
  {\bibinfo  {journal} {Nanoscale}\ }\textbf {\bibinfo {volume} {4}},\ \bibinfo
  {pages} {692} (\bibinfo {year} {2012})}\BibitemShut {NoStop}%
\bibitem [{\citenamefont {Hensen}\ \emph {et~al.}(2015)\citenamefont {Hensen},
  \citenamefont {Bernien}, \citenamefont {Dr{\'{e}}au}, \citenamefont
  {Reiserer}, \citenamefont {Kalb}, \citenamefont {Blok}, \citenamefont
  {Ruitenberg}, \citenamefont {Vermeulen}, \citenamefont {Schouten},
  \citenamefont {Abell{\'{a}}n}, \citenamefont {Amaya}, \citenamefont
  {Pruneri}, \citenamefont {Mitchell}, \citenamefont {Markham}, \citenamefont
  {Twitchen}, \citenamefont {Elkouss}, \citenamefont {Wehner}, \citenamefont
  {Taminiau},\ and\ \citenamefont {Hanson}}]{Hensen2015}%
  \BibitemOpen
  \bibfield  {author} {\bibinfo {author} {\bibfnamefont {B.}~\bibnamefont
  {Hensen}}, \bibinfo {author} {\bibfnamefont {H.}~\bibnamefont {Bernien}},
  \bibinfo {author} {\bibfnamefont {A.~E.}\ \bibnamefont {Dr{\'{e}}au}},
  \bibinfo {author} {\bibfnamefont {A.}~\bibnamefont {Reiserer}}, \bibinfo
  {author} {\bibfnamefont {N.}~\bibnamefont {Kalb}}, \bibinfo {author}
  {\bibfnamefont {M.~S.}\ \bibnamefont {Blok}}, \bibinfo {author}
  {\bibfnamefont {J.}~\bibnamefont {Ruitenberg}}, \bibinfo {author}
  {\bibfnamefont {R.~F.~L.}\ \bibnamefont {Vermeulen}}, \bibinfo {author}
  {\bibfnamefont {R.~N.}\ \bibnamefont {Schouten}}, \bibinfo {author}
  {\bibfnamefont {C.}~\bibnamefont {Abell{\'{a}}n}}, \bibinfo {author}
  {\bibfnamefont {W.}~\bibnamefont {Amaya}}, \bibinfo {author} {\bibfnamefont
  {V.}~\bibnamefont {Pruneri}}, \bibinfo {author} {\bibfnamefont {M.~W.}\
  \bibnamefont {Mitchell}}, \bibinfo {author} {\bibfnamefont {M.}~\bibnamefont
  {Markham}}, \bibinfo {author} {\bibfnamefont {D.~J.}\ \bibnamefont
  {Twitchen}}, \bibinfo {author} {\bibfnamefont {D.}~\bibnamefont {Elkouss}},
  \bibinfo {author} {\bibfnamefont {S.}~\bibnamefont {Wehner}}, \bibinfo
  {author} {\bibfnamefont {T.~H.}\ \bibnamefont {Taminiau}}, \ and\ \bibinfo
  {author} {\bibfnamefont {R.}~\bibnamefont {Hanson}},\ }\href {\doibase
  10.1038/nature15759} {\bibfield  {journal} {\bibinfo  {journal} {Nature}\
  }\textbf {\bibinfo {volume} {526}},\ \bibinfo {pages} {682} (\bibinfo {year}
  {2015})}\BibitemShut {NoStop}%
\bibitem [{\citenamefont {Fu}\ \emph {et~al.}(2013)\citenamefont {Fu},
  \citenamefont {Kuznetsov}, \citenamefont {Miroshnichenko}, \citenamefont
  {Yu},\ and\ \citenamefont {Luk'yanchuk}}]{Fu2013}%
  \BibitemOpen
  \bibfield  {author} {\bibinfo {author} {\bibfnamefont {Y.~H.}\ \bibnamefont
  {Fu}}, \bibinfo {author} {\bibfnamefont {A.~I.}\ \bibnamefont {Kuznetsov}},
  \bibinfo {author} {\bibfnamefont {A.~E.}\ \bibnamefont {Miroshnichenko}},
  \bibinfo {author} {\bibfnamefont {Y.~F.}\ \bibnamefont {Yu}}, \ and\ \bibinfo
  {author} {\bibfnamefont {B.}~\bibnamefont {Luk'yanchuk}},\ }\href {\doibase
  10.1038/ncomms2538} {\bibfield  {journal} {\bibinfo  {journal} {Nature
  Communications}\ }\textbf {\bibinfo {volume} {4}},\ \bibinfo {pages} {1527}
  (\bibinfo {year} {2013})}\BibitemShut {NoStop}%
\bibitem [{\citenamefont {Rolly}\ \emph {et~al.}(2012)\citenamefont {Rolly},
  \citenamefont {Stout},\ and\ \citenamefont {Bonod}}]{Rolly2012}%
  \BibitemOpen
  \bibfield  {author} {\bibinfo {author} {\bibfnamefont {B.}~\bibnamefont
  {Rolly}}, \bibinfo {author} {\bibfnamefont {B.}~\bibnamefont {Stout}}, \ and\
  \bibinfo {author} {\bibfnamefont {N.}~\bibnamefont {Bonod}},\ }\href
  {\doibase 10.1364/OE.20.020376} {\bibfield  {journal} {\bibinfo  {journal}
  {Optics Express}\ }\textbf {\bibinfo {volume} {20}},\ \bibinfo {pages}
  {20376} (\bibinfo {year} {2012})}\BibitemShut {NoStop}%
\bibitem [{\citenamefont {Staude}\ \emph {et~al.}(2013)\citenamefont {Staude},
  \citenamefont {Miroshnichenko}, \citenamefont {Decker}, \citenamefont
  {Fofang}, \citenamefont {Liu}, \citenamefont {Gonzales}, \citenamefont
  {Dominguez}, \citenamefont {Luk}, \citenamefont {Neshev}, \citenamefont
  {Brener},\ and\ \citenamefont {Kivshar}}]{Staude2013}%
  \BibitemOpen
  \bibfield  {author} {\bibinfo {author} {\bibfnamefont {I.}~\bibnamefont
  {Staude}}, \bibinfo {author} {\bibfnamefont {A.~E.}\ \bibnamefont
  {Miroshnichenko}}, \bibinfo {author} {\bibfnamefont {M.}~\bibnamefont
  {Decker}}, \bibinfo {author} {\bibfnamefont {N.~T.}\ \bibnamefont {Fofang}},
  \bibinfo {author} {\bibfnamefont {S.}~\bibnamefont {Liu}}, \bibinfo {author}
  {\bibfnamefont {E.}~\bibnamefont {Gonzales}}, \bibinfo {author}
  {\bibfnamefont {J.}~\bibnamefont {Dominguez}}, \bibinfo {author}
  {\bibfnamefont {T.~S.}\ \bibnamefont {Luk}}, \bibinfo {author} {\bibfnamefont
  {D.~N.}\ \bibnamefont {Neshev}}, \bibinfo {author} {\bibfnamefont
  {I.}~\bibnamefont {Brener}}, \ and\ \bibinfo {author} {\bibfnamefont
  {Y.}~\bibnamefont {Kivshar}},\ }\href {\doibase 10.1021/nn402736f} {\bibfield
   {journal} {\bibinfo  {journal} {ACS Nano}\ }\textbf {\bibinfo {volume}
  {7}},\ \bibinfo {pages} {7824} (\bibinfo {year} {2013})}\BibitemShut
  {NoStop}%
\bibitem [{\citenamefont {Benson}(2011)}]{Benson2011}%
  \BibitemOpen
  \bibfield  {author} {\bibinfo {author} {\bibfnamefont {O.}~\bibnamefont
  {Benson}},\ }\href {\doibase 10.1038/nature10610} {\bibfield  {journal}
  {\bibinfo  {journal} {Nature}\ }\textbf {\bibinfo {volume} {480}},\ \bibinfo
  {pages} {193} (\bibinfo {year} {2011})}\BibitemShut {NoStop}%
\bibitem [{\citenamefont {Kerker}\ \emph {et~al.}(1983)\citenamefont {Kerker},
  \citenamefont {Wang},\ and\ \citenamefont {Giles}}]{Kerker1983}%
  \BibitemOpen
  \bibfield  {author} {\bibinfo {author} {\bibfnamefont {M.}~\bibnamefont
  {Kerker}}, \bibinfo {author} {\bibfnamefont {D.-S.}\ \bibnamefont {Wang}}, \
  and\ \bibinfo {author} {\bibfnamefont {C.~L.}\ \bibnamefont {Giles}},\ }\href
  {\doibase 10.1364/JOSA.73.000765} {\bibfield  {journal} {\bibinfo  {journal}
  {Journal of the Optical Society of America}\ }\textbf {\bibinfo {volume}
  {73}},\ \bibinfo {pages} {765} (\bibinfo {year} {1983})}\BibitemShut
  {NoStop}%
\bibitem [{\citenamefont {Ohno}\ \emph {et~al.}(2012)\citenamefont {Ohno},
  \citenamefont {{Joseph Heremans}}, \citenamefont {Bassett}, \citenamefont
  {Myers}, \citenamefont {Toyli}, \citenamefont {{Bleszynski Jayich}},
  \citenamefont {Palmstr{\o}m},\ and\ \citenamefont {Awschalom}}]{Ohno2012}%
  \BibitemOpen
  \bibfield  {author} {\bibinfo {author} {\bibfnamefont {K.}~\bibnamefont
  {Ohno}}, \bibinfo {author} {\bibfnamefont {F.}~\bibnamefont {{Joseph
  Heremans}}}, \bibinfo {author} {\bibfnamefont {L.~C.}\ \bibnamefont
  {Bassett}}, \bibinfo {author} {\bibfnamefont {B.~A.}\ \bibnamefont {Myers}},
  \bibinfo {author} {\bibfnamefont {D.~M.}\ \bibnamefont {Toyli}}, \bibinfo
  {author} {\bibfnamefont {A.~C.}\ \bibnamefont {{Bleszynski Jayich}}},
  \bibinfo {author} {\bibfnamefont {C.~J.}\ \bibnamefont {Palmstr{\o}m}}, \
  and\ \bibinfo {author} {\bibfnamefont {D.~D.}\ \bibnamefont {Awschalom}},\
  }\href {\doibase 10.1063/1.4748280} {\bibfield  {journal} {\bibinfo
  {journal} {Applied Physics Letters}\ }\textbf {\bibinfo {volume} {101}},\
  \bibinfo {pages} {082413} (\bibinfo {year} {2012})}\BibitemShut {NoStop}%
\bibitem [{\citenamefont {Michl}\ \emph {et~al.}(2014)\citenamefont {Michl},
  \citenamefont {Teraji}, \citenamefont {Zaiser}, \citenamefont {Jakobi},
  \citenamefont {Waldherr}, \citenamefont {Dolde}, \citenamefont {Neumann},
  \citenamefont {Doherty}, \citenamefont {Manson}, \citenamefont {Isoya},\ and\
  \citenamefont {Wrachtrup}}]{Michl2014}%
  \BibitemOpen
  \bibfield  {author} {\bibinfo {author} {\bibfnamefont {J.}~\bibnamefont
  {Michl}}, \bibinfo {author} {\bibfnamefont {T.}~\bibnamefont {Teraji}},
  \bibinfo {author} {\bibfnamefont {S.}~\bibnamefont {Zaiser}}, \bibinfo
  {author} {\bibfnamefont {I.}~\bibnamefont {Jakobi}}, \bibinfo {author}
  {\bibfnamefont {G.}~\bibnamefont {Waldherr}}, \bibinfo {author}
  {\bibfnamefont {F.}~\bibnamefont {Dolde}}, \bibinfo {author} {\bibfnamefont
  {P.}~\bibnamefont {Neumann}}, \bibinfo {author} {\bibfnamefont {M.~W.}\
  \bibnamefont {Doherty}}, \bibinfo {author} {\bibfnamefont {N.~B.}\
  \bibnamefont {Manson}}, \bibinfo {author} {\bibfnamefont {J.}~\bibnamefont
  {Isoya}}, \ and\ \bibinfo {author} {\bibfnamefont {J.}~\bibnamefont
  {Wrachtrup}},\ }\href {\doibase 10.1063/1.4868128} {\bibfield  {journal}
  {\bibinfo  {journal} {Applied Physics Letters}\ }\textbf {\bibinfo {volume}
  {104}},\ \bibinfo {pages} {102407} (\bibinfo {year} {2014})}\BibitemShut
  {NoStop}%
\bibitem [{\citenamefont {Alaee}\ \emph {et~al.}(2015)\citenamefont {Alaee},
  \citenamefont {Filter}, \citenamefont {Lehr}, \citenamefont {Lederer},\ and\
  \citenamefont {Rockstuhl}}]{Alaee2015}%
  \BibitemOpen
  \bibfield  {author} {\bibinfo {author} {\bibfnamefont {R.}~\bibnamefont
  {Alaee}}, \bibinfo {author} {\bibfnamefont {R.}~\bibnamefont {Filter}},
  \bibinfo {author} {\bibfnamefont {D.}~\bibnamefont {Lehr}}, \bibinfo {author}
  {\bibfnamefont {F.}~\bibnamefont {Lederer}}, \ and\ \bibinfo {author}
  {\bibfnamefont {C.}~\bibnamefont {Rockstuhl}},\ }\href {\doibase
  10.1364/OL.40.002645} {\bibfield  {journal} {\bibinfo  {journal} {Optics
  Letters}\ }\textbf {\bibinfo {volume} {40}},\ \bibinfo {pages} {2645}
  (\bibinfo {year} {2015})}\BibitemShut {NoStop}%
\bibitem [{\citenamefont {Nelder}\ and\ \citenamefont
  {Mead}(1965)}]{Nelder1965}%
  \BibitemOpen
  \bibfield  {author} {\bibinfo {author} {\bibfnamefont {J.~A.}\ \bibnamefont
  {Nelder}}\ and\ \bibinfo {author} {\bibfnamefont {R.}~\bibnamefont {Mead}},\
  }\href {\doibase 10.1093/comjnl/7.4.308} {\bibfield  {journal} {\bibinfo
  {journal} {The Computer Journal}\ }\textbf {\bibinfo {volume} {7}},\ \bibinfo
  {pages} {308} (\bibinfo {year} {1965})}\BibitemShut {NoStop}%
\bibitem [{\citenamefont {Person}\ \emph {et~al.}(2013)\citenamefont {Person},
  \citenamefont {Jain}, \citenamefont {Lapin}, \citenamefont {S{\'{a}}enz},
  \citenamefont {Wicks},\ and\ \citenamefont {Novotny}}]{Person2013}%
  \BibitemOpen
  \bibfield  {author} {\bibinfo {author} {\bibfnamefont {S.}~\bibnamefont
  {Person}}, \bibinfo {author} {\bibfnamefont {M.}~\bibnamefont {Jain}},
  \bibinfo {author} {\bibfnamefont {Z.}~\bibnamefont {Lapin}}, \bibinfo
  {author} {\bibfnamefont {J.~J.}\ \bibnamefont {S{\'{a}}enz}}, \bibinfo
  {author} {\bibfnamefont {G.}~\bibnamefont {Wicks}}, \ and\ \bibinfo {author}
  {\bibfnamefont {L.}~\bibnamefont {Novotny}},\ }\href {\doibase
  10.1021/nl4005018} {\bibfield  {journal} {\bibinfo  {journal} {Nano Letters}\
  }\textbf {\bibinfo {volume} {13}},\ \bibinfo {pages} {1806} (\bibinfo {year}
  {2013})}\BibitemShut {NoStop}%
\bibitem [{\citenamefont {Livneh}\ \emph {et~al.}(2016)\citenamefont {Livneh},
  \citenamefont {Harats}, \citenamefont {Istrati}, \citenamefont {Eisenberg},\
  and\ \citenamefont {Rapaport}}]{Livneh2016}%
  \BibitemOpen
  \bibfield  {author} {\bibinfo {author} {\bibfnamefont {N.}~\bibnamefont
  {Livneh}}, \bibinfo {author} {\bibfnamefont {M.~G.}\ \bibnamefont {Harats}},
  \bibinfo {author} {\bibfnamefont {D.}~\bibnamefont {Istrati}}, \bibinfo
  {author} {\bibfnamefont {H.~S.}\ \bibnamefont {Eisenberg}}, \ and\ \bibinfo
  {author} {\bibfnamefont {R.}~\bibnamefont {Rapaport}},\ }\href {\doibase
  10.1021/acs.nanolett.6b00082} {\bibfield  {journal} {\bibinfo  {journal}
  {Nano Letters}\ }\textbf {\bibinfo {volume} {16}},\ \bibinfo {pages} {2527}
  (\bibinfo {year} {2016})}\BibitemShut {NoStop}%
\bibitem [{\citenamefont {Li}\ \emph {et~al.}(2015)\citenamefont {Li},
  \citenamefont {Chen}, \citenamefont {Zheng}, \citenamefont {Mouradian},
  \citenamefont {Dolde}, \citenamefont {Schr{\"{o}}der}, \citenamefont
  {Karaveli}, \citenamefont {Markham}, \citenamefont {Twitchen},\ and\
  \citenamefont {Englund}}]{Li2015}%
  \BibitemOpen
  \bibfield  {author} {\bibinfo {author} {\bibfnamefont {L.}~\bibnamefont
  {Li}}, \bibinfo {author} {\bibfnamefont {E.~H.}\ \bibnamefont {Chen}},
  \bibinfo {author} {\bibfnamefont {J.}~\bibnamefont {Zheng}}, \bibinfo
  {author} {\bibfnamefont {S.~L.}\ \bibnamefont {Mouradian}}, \bibinfo {author}
  {\bibfnamefont {F.}~\bibnamefont {Dolde}}, \bibinfo {author} {\bibfnamefont
  {T.}~\bibnamefont {Schr{\"{o}}der}}, \bibinfo {author} {\bibfnamefont
  {S.}~\bibnamefont {Karaveli}}, \bibinfo {author} {\bibfnamefont {M.~L.}\
  \bibnamefont {Markham}}, \bibinfo {author} {\bibfnamefont {D.~J.}\
  \bibnamefont {Twitchen}}, \ and\ \bibinfo {author} {\bibfnamefont
  {D.}~\bibnamefont {Englund}},\ }\href {\doibase 10.1021/nl503451j} {\bibfield
   {journal} {\bibinfo  {journal} {Nano Letters}\ }\textbf {\bibinfo {volume}
  {15}},\ \bibinfo {pages} {1493} (\bibinfo {year} {2015})}\BibitemShut
  {NoStop}%
\bibitem [{\citenamefont {Moroz}(2010)}]{Moroz2010}%
  \BibitemOpen
  \bibfield  {author} {\bibinfo {author} {\bibfnamefont {A.}~\bibnamefont
  {Moroz}},\ }\href {\doibase 10.1016/j.optcom.2010.01.061} {\bibfield
  {journal} {\bibinfo  {journal} {Optics Communications}\ }\textbf {\bibinfo
  {volume} {283}},\ \bibinfo {pages} {2277} (\bibinfo {year}
  {2010})}\BibitemShut {NoStop}%
\end{thebibliography}%


\begin{thebibliography}{3}%
\makeatletter
\providecommand \@ifxundefined [1]{%
 \@ifx{#1\undefined}
}%
\providecommand \@ifnum [1]{%
 \ifnum #1\expandafter \@firstoftwo
 \else \expandafter \@secondoftwo
 \fi
}%
\providecommand \@ifx [1]{%
 \ifx #1\expandafter \@firstoftwo
 \else \expandafter \@secondoftwo
 \fi
}%
\providecommand \natexlab [1]{#1}%
\providecommand \enquote  [1]{``#1''}%
\providecommand \bibnamefont  [1]{#1}%
\providecommand \bibfnamefont [1]{#1}%
\providecommand \citenamefont [1]{#1}%
\providecommand \href@noop [0]{\@secondoftwo}%
\providecommand \href [0]{\begingroup \@sanitize@url \@href}%
\providecommand \@href[1]{\@@startlink{#1}\@@href}%
\providecommand \@@href[1]{\endgroup#1\@@endlink}%
\providecommand \@sanitize@url [0]{\catcode `\\12\catcode `\$12\catcode
  `\&12\catcode `\#12\catcode `\^12\catcode `\_12\catcode `\%12\relax}%
\providecommand \@@startlink[1]{}%
\providecommand \@@endlink[0]{}%
\providecommand \url  [0]{\begingroup\@sanitize@url \@url }%
\providecommand \@url [1]{\endgroup\@href {#1}{\urlprefix }}%
\providecommand \urlprefix  [0]{URL }%
\providecommand \Eprint [0]{\href }%
\providecommand \doibase [0]{http://dx.doi.org/}%
\providecommand \selectlanguage [0]{\@gobble}%
\providecommand \bibinfo  [0]{\@secondoftwo}%
\providecommand \bibfield  [0]{\@secondoftwo}%
\providecommand \translation [1]{[#1]}%
\providecommand \BibitemOpen [0]{}%
\providecommand \bibitemStop [0]{}%
\providecommand \bibitemNoStop [0]{.\EOS\space}%
\providecommand \EOS [0]{\spacefactor3000\relax}%
\providecommand \BibitemShut  [1]{\csname bibitem#1\endcsname}%
\let\auto@bib@innerbib\@empty
\bibitem [{\citenamefont {Burger}\ \emph {et~al.}(2009)\citenamefont {Burger},
  \citenamefont {Kleemann}, \citenamefont {Zschiedrich},\ and\ \citenamefont
  {Schmidt}}]{Burger2009}%
  \BibitemOpen
  \bibfield  {author} {\bibinfo {author} {\bibfnamefont {S.}~\bibnamefont
  {Burger}}, \bibinfo {author} {\bibfnamefont {B.~H.}\ \bibnamefont
  {Kleemann}}, \bibinfo {author} {\bibfnamefont {L.}~\bibnamefont
  {Zschiedrich}}, \ and\ \bibinfo {author} {\bibfnamefont {F.}~\bibnamefont
  {Schmidt}}\ }(\bibinfo {year} {2009})\ p.\ \bibinfo {pages}
  {736621}\BibitemShut {NoStop}%
\bibitem [{\citenamefont {Peter}(1923)}]{Peter1923}%
  \BibitemOpen
  \bibfield  {author} {\bibinfo {author} {\bibfnamefont {F.}~\bibnamefont
  {Peter}},\ }\href {\doibase 10.1007/BF01330487} {\bibfield  {journal}
  {\bibinfo  {journal} {Zeitschrift f{\"{u}}r Physik}\ }\textbf {\bibinfo
  {volume} {15}},\ \bibinfo {pages} {358} (\bibinfo {year} {1923})}\BibitemShut
  {NoStop}%
\bibitem [{\citenamefont {Novotny}\ and\ \citenamefont
  {Hecht}(2006)}]{Novotny2006}%
  \BibitemOpen
  \bibfield  {author} {\bibinfo {author} {\bibfnamefont {L.}~\bibnamefont
  {Novotny}}\ and\ \bibinfo {author} {\bibfnamefont {B.}~\bibnamefont
  {Hecht}},\ }\href {\doibase 10.1017/CBO9780511813535} {\emph {\bibinfo
  {title} {{Principles of Nano-Optics}}}}\ (\bibinfo  {publisher} {Cambridge
  University Press},\ \bibinfo {address} {Cambridge},\ \bibinfo {year}
  {2006})\BibitemShut {NoStop}%
\end{thebibliography}%
\end{document}



\title{Supplementary material: A compact and simple photonic nanoantenna for efficient photon extraction from nitrogen vacancy centers in bulk diamond}

\author{Niko Nikolay}
\author{Günter Kewes}%
\author{Oliver Benson}%
\affiliation{%
 Institut für Physik, Humboldt-Universität zu Berlin,\\
 Newtonstraße 15, D-12489 Berlin, Germany
}%

\maketitle

\section{Simulation setup}
\label{sec:simulation}

Simulations were done with JCMwave, a full 3D frequency-domain finite-element solver. In all simulations the cylindrical symmetry was exploited \cite{Burger2009}. The frequency dependent refractive index of Diamond was taken from Ref. \cite{Peter1923}. All simulations were done with a horizontally polarized dipole. Collection efficiencies were calculated by integrating the intensity directed into the far field within angles corresponding to a certain NA centered in the lower plane, normalized by the whole intensity emitted by the dipole. Details are discussed in another section.

\section{Wavelength sampling}
\label{sec:wvlsampling}
\begin{figure}
 \centering
 \includegraphics[width=\textwidth]{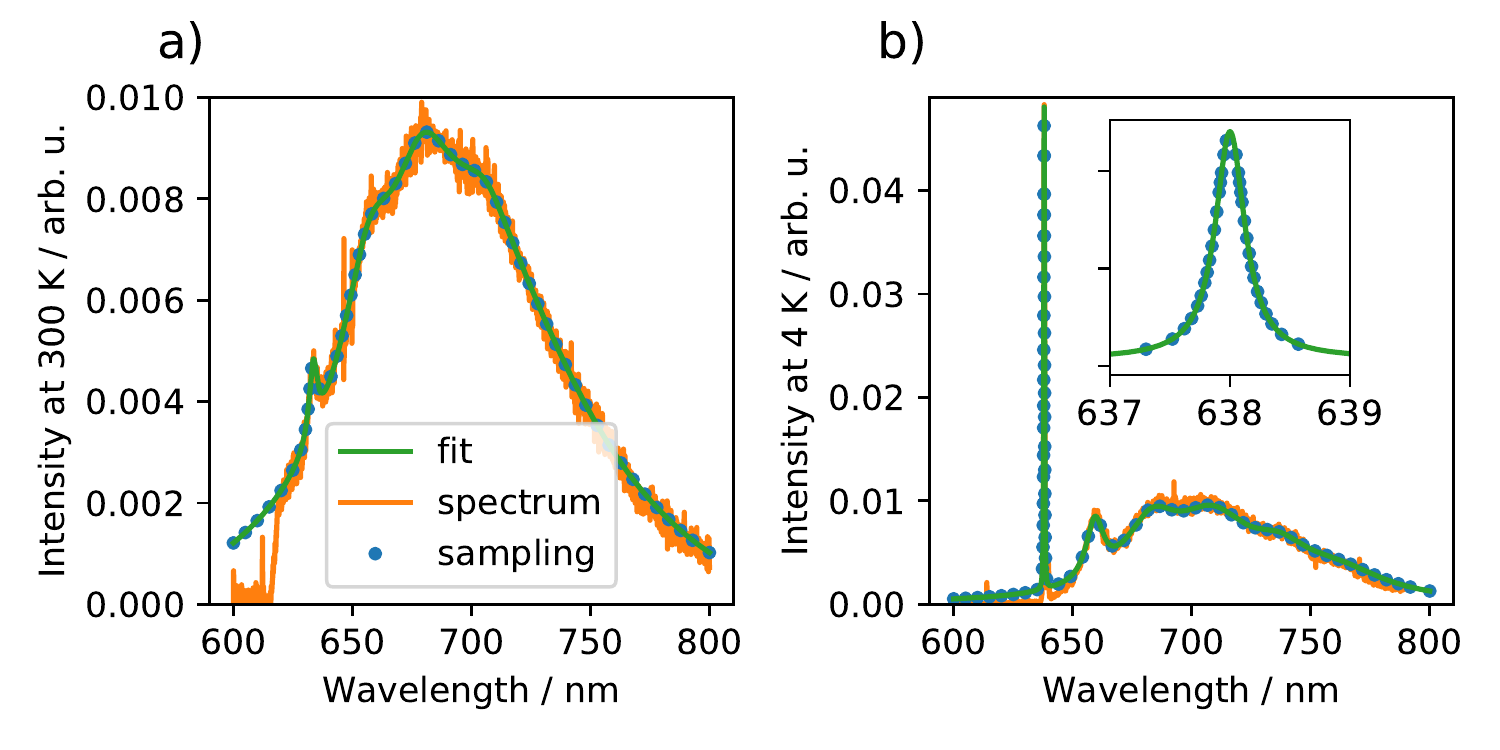}
 \caption{\textit{Numerical sampling of spectra at cryogenic and room temperature.} The green line show a fit on orange data sets of measured NV spectra taken at \SI{300}{\kelvin} (a) and at \SI{5}{\kelvin} (b). Blue dots denote sampling points, on which numerical simulations evaluated the antenna performance. The inset shows a zoom into the zero phonon line with the increased sampling density.}
 \label{fig:spectra}
\end{figure}

To ensure that all spectral features of the coupled system (NV and antenna) are properly considered, each system was analyzed separately in advance. We found, that an antenna structure optimized for a single frequency shows resonances with a FWHM of $\sim\SI{15}{\nano\meter}$. Thus, we used \SI{5}{\nano\meter} steps to ensure that these features are fully taken into account. However, the ZPL of the NV is even sharper, especially at cryogenic temperatures. Therefore, an adaptive sampling was done so that the uncertainty of the sampled area (calculated using the trapezoidal rule) is below \SI{1}{\percent}. The final wavelength sampling for two spectra taken at room temperature and at cryogenic temperatures are shown in \autoref{fig:spectra} indicated by blue dots.

\section{Relative reduction of the lifetime and collection efficiency}
\label{sec:computation}

A postprocess from JCMwave gives the amount of emitted power $p$ from a dipole placed within the simulation domain. The lifetime shortening is given in this situation by \cite{Novotny2006}:
\begin{equation}
\frac{\tau'}{\tau} = \frac{p}{p'}
\end{equation}
For realistic spectrally broad emitters, it is not sufficient to consider a lifetime change at just a single wavelength, but one has to average over the entire spectrum ($\Delta\lambda=\lambda_\text{full}$):
\begin{equation}\label{eqn:power}
p_\text{total} = \int\limits_{\lambda_\text{full}} p(\lambda)\,\rho(\lambda)\,\mathbf{d}\lambda
\end{equation}

Using a near to far field transformation from JCMwave, the emitted electric field strength into a certain angle is extracted. From that we can calculate the directed intensity:

\begin{equation}
I'(\theta,\phi) = \frac{1}{2}\,\text{c}\,n\,\epsilon_0\,\left|E(\theta,\phi)\right|^2
\end{equation}

Note that this is the intensity directed into the substrate. The reflection from the diamond-air/oil surface has been included by decomposing the far field into p- and s-polarized light, calculating the transmission coefficients with the Fresnel equations for each polarization and finally weight the intensity $I'$ with the resulting transmission coefficient.

Directivies are given by
\begin{equation}
D(\theta,\phi) = \frac{I(\theta,\phi)}{4\,\pi\,p}
\end{equation}
with the transmitted intensity $I(\theta,\phi)=T(\theta,\phi)\cdot I'(\theta,\phi)$. Applying this directivity to the whole emitter spectrum can be done analog to \autoref{eqn:power}.

\section{Reference}
\label{sec:sweetspot}

\begin{figure}
 \centering
 \includegraphics[width=\textwidth]{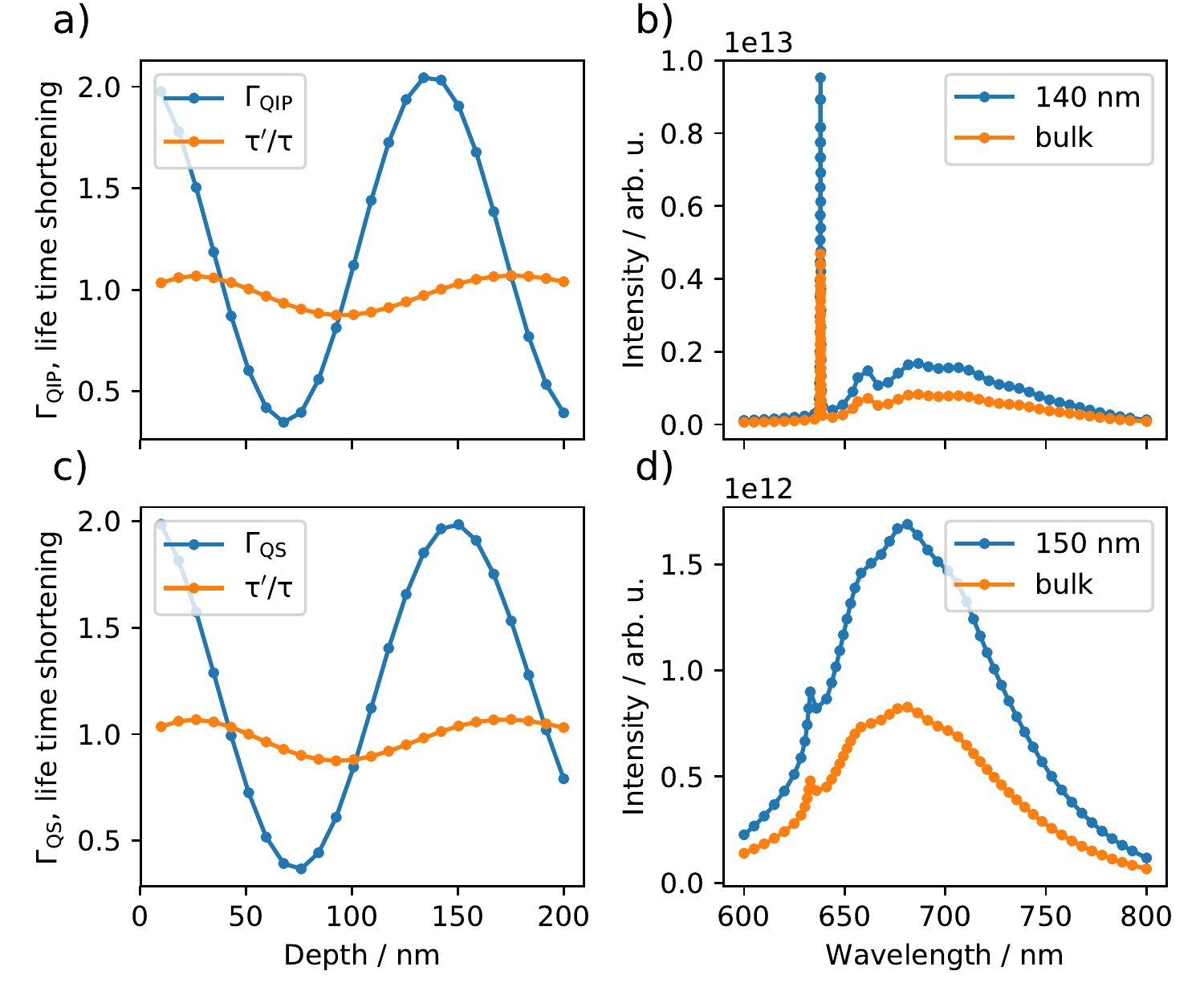}
 \caption{\textit{$\varGamma_\text{QIP}$ of a layered system relative to a dipole in a bulk diamond.} In a) and c) the dipole depth below the diamond-air interface is sweeped while the lifetime shortening (green) and the a) $\Gamma_\text{QIP}$ and b) $\Gamma_\text{QS}$ (denoted in blue) was evaluated. Altered spectra calculated at a depth of \SI{140}{\nano\meter} (blue) and for bulk diamond (orange) are shown in b) and d) for each $\Gamma$ respectively.}
 \label{fig:sweetspot}
\end{figure}

To get a measure of how much better the proposed antenna design will be, a reference has to be determined. As a first step, the emission properties of a dipole within an infinite bulk diamond was calculated. In a realistic experiment, the NV is often in close vicinity to the bulk diamond surface, while the collection is done from the substrate side. Even in such a configuration, a sweet spot which maximizes $\Gamma_\text{QIP}$ or $\Gamma_\text{QS}$ exists. A depth scan was done and compared to the bulk diamond, results are shown in \autoref{fig:sweetspot}. With the restriction of having a depth of at least \SI{5}{\nano\meter}, an optimal depth of \SI{140}{\nano\meter} for  $\Gamma'_\text{QIP}$ (a) and \SI{150}{\nano\meter} for $\Gamma'_\text{QS}$ (c) was found. Altered spectra for the optimal and for the worst position are shown in \autoref{fig:sweetspot} b) and d) respectively.

\section{Optimization}
\label{sec:optimization}

\begin{figure}
 \centering
 \includegraphics[width=\textwidth]{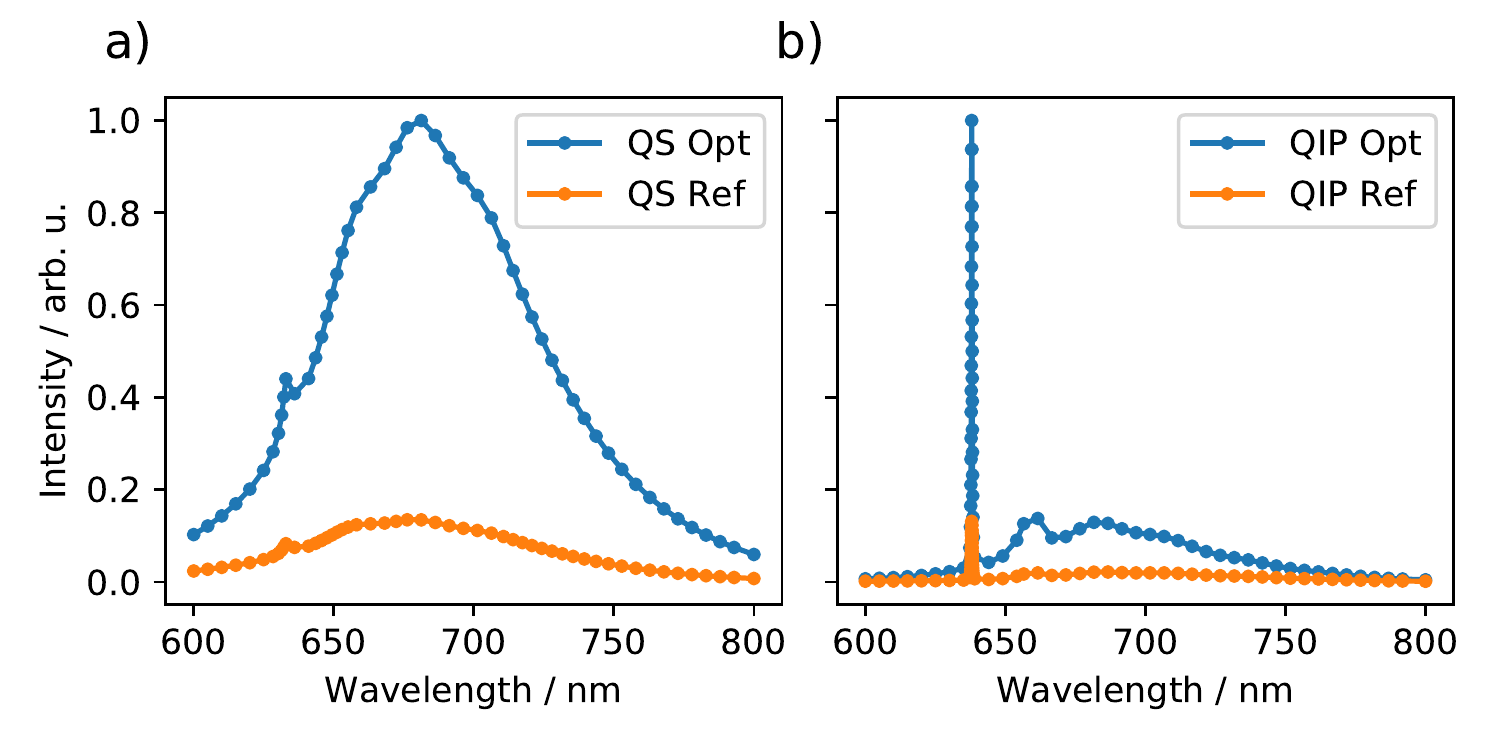}
 \caption{\textit{Antenna modified spectra.} Altered spectra due to antenna with optimal parameter sets and the unperturbed reference spectrum for $\Gamma_\text{QS}$ a) and $\Gamma_\text{QIP}$ b) are shown.}
 \label{fig:spectral_response}
\end{figure}

The simplex algorithm was used to find maxima. To optimize $\Gamma_\text{QIP}$, the dipole depth was limited to a minimal value of \SI{50}{\nano\meter}, to have long decoherence times, and a maximal value of \SI{200}{\nano\meter} to save computation time. We also found a second local maximum apart from the global maximum mentioned in the main paper with the following parameters: radius \SI{140}{\nano\meter}, height \SI{260}{\nano\meter}, dipole depth \SI{139}{\nano\meter} and $\Gamma_\text{QIP}$ 4.5. $\Gamma_\text{QS}$ was optimized in the same way but with the restriction of having a depth of at least \SI{5}{\nano\meter}. Spectra stemming from an optimal structure are shown in \autoref{fig:spectral_response} denoted by blue dots. For comparison the spectra used as reference are again shown, denoted by orange dots.

\section{Parameters for optimal collection efficiency and upward directivity}
\label{sec:opt_params}

For other applications, like in quantum communication, an optimal collection efficiency rather than lifetime shortening might be desired. $\Gamma$ is then given by

\begin{equation}
\Gamma = \frac{\int\limits_{\text{NV}}\rho(\lambda)\,\eta_C(NA,\lambda)\,\mathbf{d}\lambda}{\int\limits_{\text{NV}}\rho(\lambda)\,\eta'_C(NA,\lambda)\,\mathbf{d}\lambda}
\end{equation}

\begin{table}[ht]
 \caption{\textit{Optimized values.} Optimal geometric dimensions for sensing and a upward directivity are shown. Geometric dimensions are given in nm.}
 \begin{ruledtabular}
  \begin{tabular}{c c c c c c c c}
   & height& radius & depth & $\Gamma^{0.4}$ & $\Gamma^{0.6}$ & $\Gamma^{0.9}$ & $\Gamma^{1.4, \text{oil}}$\\
   \hline
   Sensing	& $461$ & $397$ & $5$ & $4.9$ & $3.7$ & $4.2$ & $6.2$ \\
   Upward	& $588$ & $301$ & $342$ & $44$ & $17$ & $7$ & $-$ \\
  \end{tabular}
 \end{ruledtabular}
 \label{tab:results}
\end{table}

\section{Convergence test}
\label{sec:convergence}

\begin{figure}
 \centering
 \includegraphics[width=0.5\textwidth]{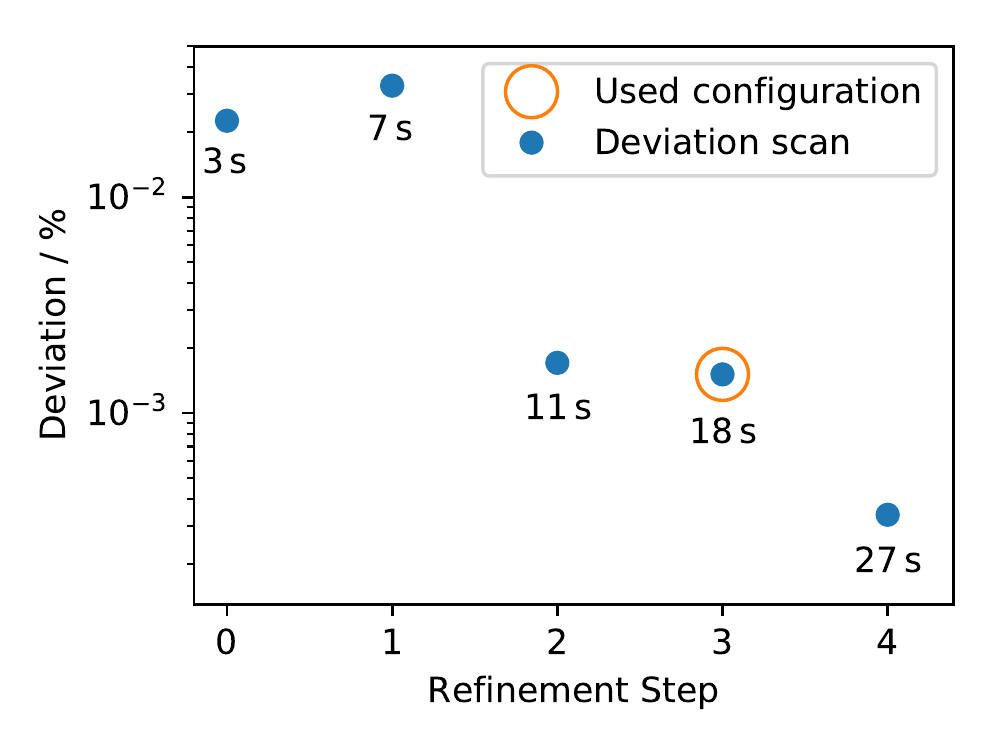}
 \caption{\textit{Convergence test and computation times.} The power guided into NA 0.9 extracted from test simulations compared to a very precise reference simulation is shown. Numbers below points denote the needed CPU time (single thread) on a Xeon E5-2637 v3 @ 3.5 GHz CPU. The used configuration is denoted by an orange circle.}
 \label{fig:convergence}
\end{figure}

To prove for convergence of the Maxwell solver JCMwave, a reference with an extremely fine starting grid, 5 steps of adaptive refinement and large simulation domains was calculated. From this simulation the far field intensity directed into a NA 0.9 was extracted, since this is a crucial number for all the simulations done here. Subsequently, this result was compared to simulations done with a coarse starting mesh and a smaller computation domain depending on the number of refinement steps. Results are shown in \autoref{fig:convergence}. As a trade off between accuracy and time consumption, three refinement steps were used to compute all the other data presented in this publication.

\bibliography{references}